\documentclass{naturxiv}
\usepackage[pdftex]{graphicx}
\usepackage[english]{babel}  %
\usepackage{amssymb}         %
\usepackage[fleqn]{amsmath}         %
\usepackage{enumerate}
\usepackage[T1]{fontenc}
\usepackage{makeidx}
\usepackage[normalem]{ulem}
\usepackage{color}
\usepackage{bm}%
\usepackage{hyperref}   %
\usepackage{booktabs} %
\usepackage{tabularx} %
\usepackage{wrapfig}
\usepackage{pdfpages}
\title{Discovering the laws of urbanisation}
\author{Filippo Simini$^{1}$, \& Charlotte R. James$^1$}
\begin{document}
\maketitle
\begin{small}
\begin{affiliations}
 \item Department of Engineering Mathematics, University of Bristol, Merchant Venturers Building, Woodland Road, BS8 1UB, Bristol, UK.
\end{affiliations}
\end{small}
\begin{abstract}
In 2012 the world's population exceeded 7 billion, and since 2008 the number of individuals living in urban areas has surpassed that of rural areas. This is the result of an overall increase of life expectancy in many countries that has caused an unprecedented growth of the world's total population during recent decades, combined with a net migration flow from rural villages to urban agglomerations\cite{RefWorks:296}. 
While it is clear that the rate of natural increase and migration flows are the driving forces shaping the spatial distribution of population, a general consensus on the mechanisms that characterise the urbanisation process is still lacking 
\cite{RefWorks:173,RefWorks:187,RefWorks:167,RefWorks:307,RefWorks:306,RefWorks:304}. 
Here we present two fundamental laws of urbanisation that are quantitatively supported by empirical evidence: 1) the number of cities in a country is proportional to the country's total population, irrespective of the country's area, and 2) the average distance between cities scales as the inverse of the square root of the country's population density. 
We study the spatio-temporal evolution of population considering two classes of models, Gravity and Intervening Opportunities, to estimate migration flows and show that they produce different spatial patterns of cities. 
\end{abstract}

\begin{figure}
\begin{center}
\includegraphics[type=pdf,ext=.pdf,read=.pdf,width=170mm]{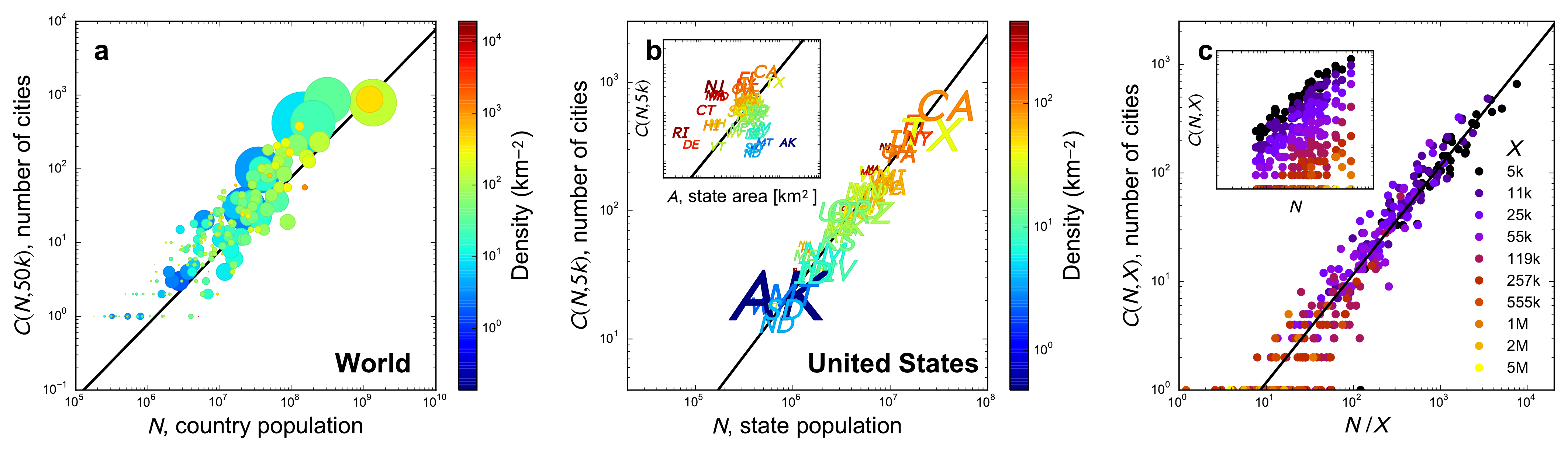}
\end{center}
\caption{{
The first law of urbanisation. 
{\bf a},  The number of cities with more than 50,000 inhabitants, $C(N, 50k)$ ($y$ value), as a function of the population $N$ ($x$ value), area $A$ (circle size), and population density $\rho$ (color), for 210 countries. 
The correlation coefficients $\text{corr}(C,N) = 0.84$, $\text{corr}(C,A) = 0.64$, and $\text{corr}(C,\rho) = -0.04$ indicate a strong linear dependence between $C$ and $N$. 
The weaker correlation between a country's area and the number of cities is due to an indirect dependence of area and total population, i.e. larger countries tend to be more populated. 
{\bf b}, The number of cities with more than 5,000 inhabitants in the Unites States is proportional to the state's population, $\text{corr}(C,N) = 0.95$. The correlations with area ($0.04$, see inset) and population density ($-0.08$) are negligible, as illustrated by the following pairs of states with similar area or density and very different number of cities: Alaska ("AK": $A=1.5$M km$^2$, $C(5k)=22$) vs Texas  ("TX": $A=0.7$M km$^2$, $C(5k)=392$), and Rhode Island ("RI": $\rho=393$ km$^{-2}$, $C(5k)=35$) vs New Jersey ("NJ": $\rho=467$ km$^{-2}$, $C(5k)=316$). 
{\bf c},  Combining the first law of urbanisation with Zipf's law it is possible to estimate the number of cities with more than $X$ inhabitants in a country with population $N$ as $C(N,X) \sim N/X$. 
As a consequence, the scattered cloud of points resulting when plotting $C(N,X)$ against $N$ for various $X$'s in the range $5 \cdot 10^3 - 5 \cdot 10^6$ (inset) collapses on a straight line when $C(N,X)$ is plotted against the ratio $N/X$. 
} \label{fig:1}}
\end{figure}

\begin{figure}
\begin{center}
\includegraphics[type=pdf,ext=.pdf,read=.pdf,width=170mm]{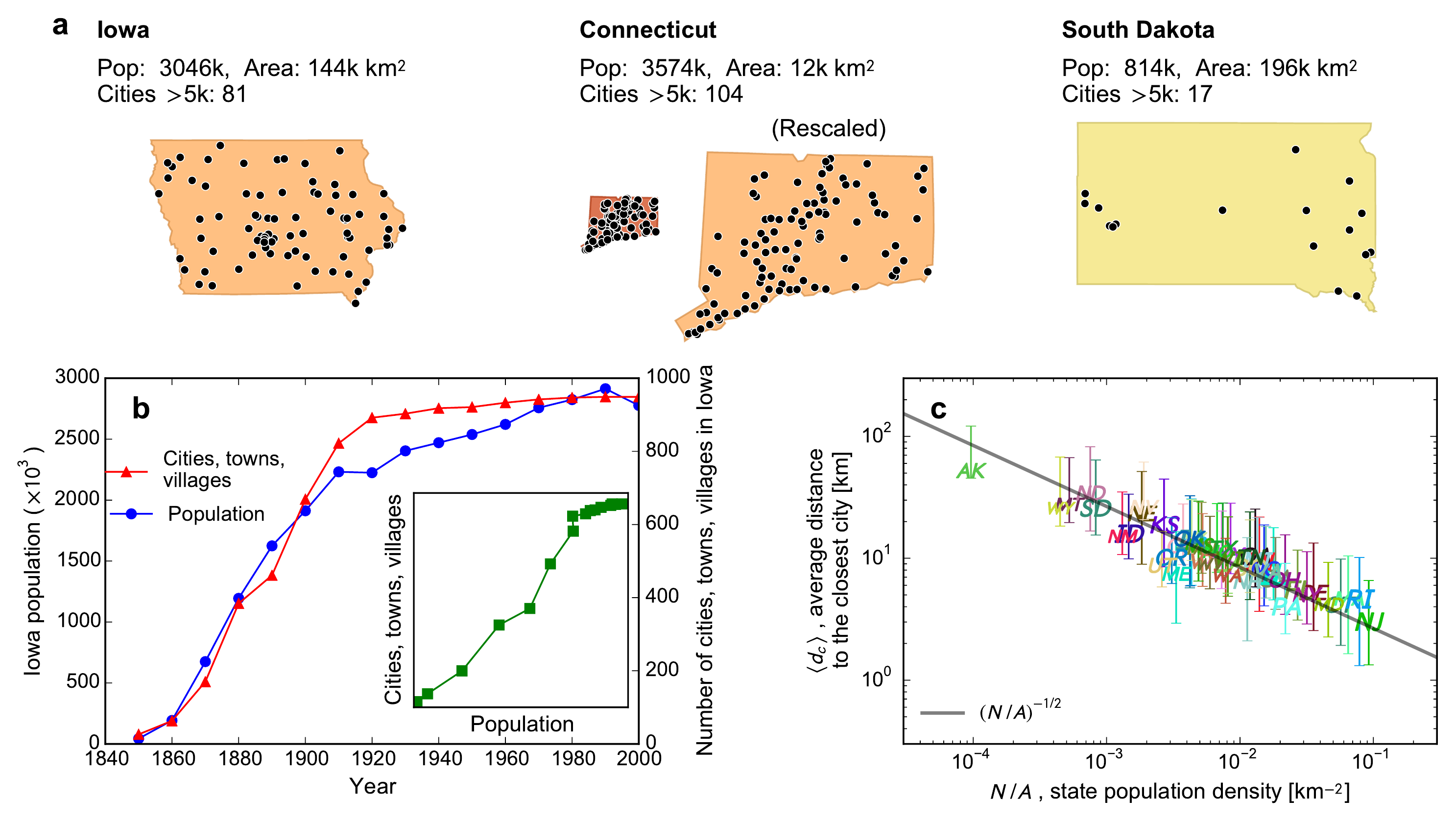}
\end{center}
\caption{{
{\bf a},  Illustration of the relationships between total population, number of cities, and their average distance
in Iowa, Connecticut, and South Dakota.
In agreement with the first law of urbanisation, Eq.\ref{eq:1st_law}, Iowa and Connecticut have similar populations and a similar number of cities with more than 5,000 inhabitants, despite Connecticut having one-twelfth the area of Iowa; South Dakota instead has roughly one-fourth of both the population and number of cities of Iowa and Connecticut, despite having a slightly larger area than Iowa. 
In agreement with the second law of urbanisation, Eq.\ref{eq:2nd_law}, cities in Connecticut are closer than cities in Iowa because of the higher population density in Connecticut. By rescaling distances such that Connecticut's area becomes equal to Iowa's area, the two states would have the same population density and consequently the same average distance between cities.
{\bf b},  Historical records of the number of incorporated places ($C$, red triangles) and the state population ($N$, blue circles) in Iowa from 1850 to 2000 (source: State library of Iowa, state data center). 
The similar growth rates of $C$ and $N$ entail the validity of the first law of urbanisation $C \sim N$ during the 150-year period (inset). 
In a country with a uniformly growing population the first law of urbanisation implies the stationarity of Zipf's law and vice versa: 
If the population of each city doubles in a time $\Delta t$, the number of cities with more than $X$ inhabitants at time $t+\Delta t$ is equal to the number of cities that had more than $X/2$ inhabitants at time $t$, $C(2N) P_>(X, t + \Delta t) = C(N)  P_>(X/2, t)$, then the Zipf law is the stationary distribution of city sizes $P_>(X, t + \Delta t) = P_>(X, t) = 1/X$, if the first law of urbanisation holds $C(2N) = 2 C(N) \Rightarrow C(N) \propto N$. 
{\bf c},  The second law of urbanisation for the United States. The average distance to the closest city scales as the inverse of the square root of the state's population density (here all cities with more than 5,000 inhabitants are considered). The asymmetric error bars denote the standard deviations above and below the average.  
} \label{fig:2}}
\end{figure}

\begin{figure}
\begin{center}
\includegraphics[type=pdf,ext=.pdf,read=.pdf,width=170mm]{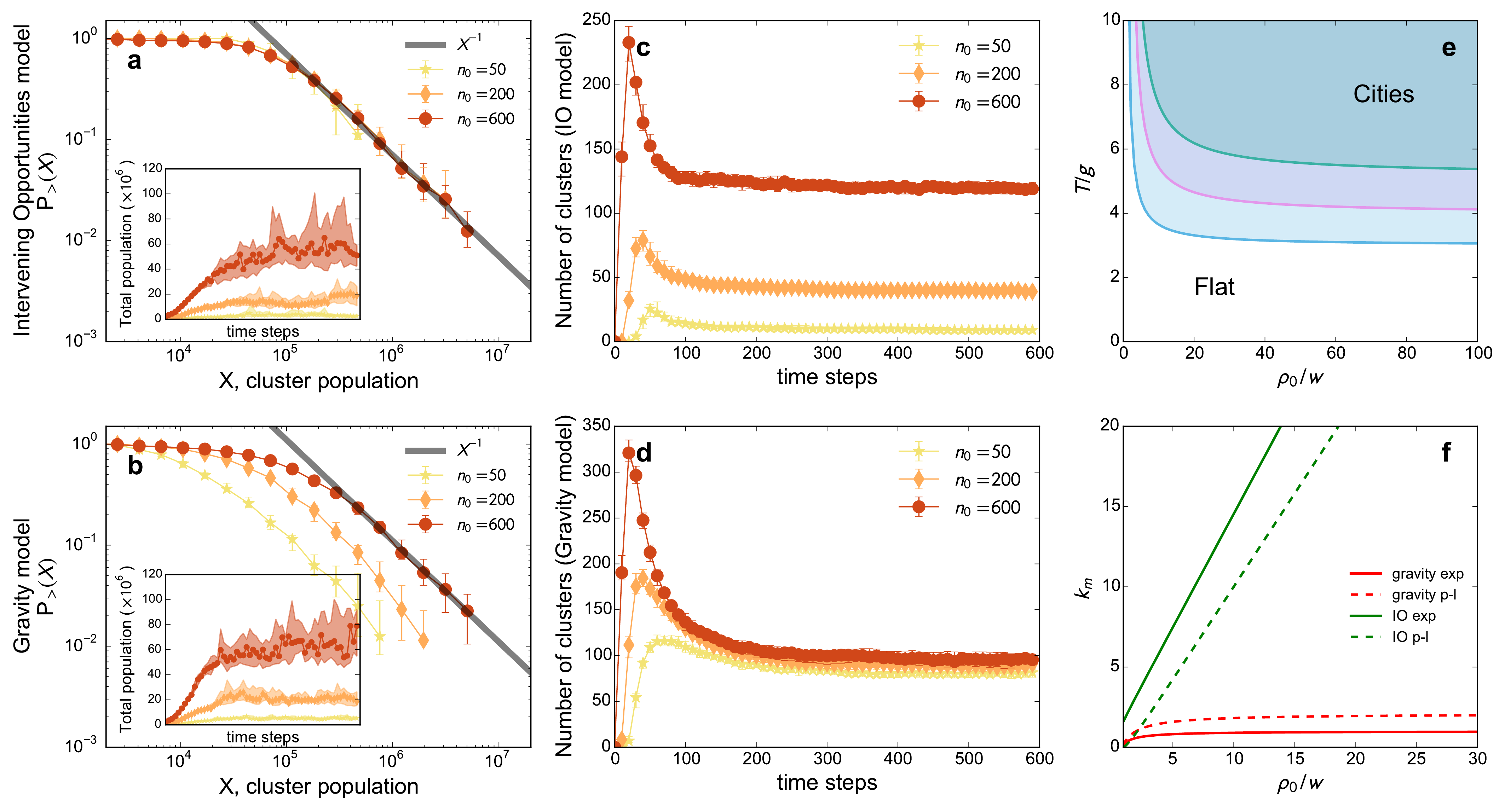}
\end{center}
\caption{{
{\bf a-b}, 
Main panels, counter cumulative distributions of clusters' populations, $X$, at the end of the numerical simulations using Intervening Opportunities (a) and Gravity (b) models of migration, for various values of minimum population $n_0 = 50, 200, 600$ (see Supplementary Information for simulation details). 
The grey line is a guide for the eye depicting Zipf's law, $P_>(X) \sim 1/X$. 
Symbols at each $X$ denote the average fraction of clusters with population larger than $X$ over 16 realizations of the numerical simulation, and error bars represent the 25th and 75th percentiles. 
Insets, total population as a function of the simulation time steps. 
Lines with symbols denote the mean total population whereas each shaded area represents the region between the 25th and 75th percentiles.
Simulations with larger $n_0$ have a larger total population at all time steps for both models, implying that the total population increases with $n_0$.
{\bf c-d}, 
Time evolution of the number of clusters obtained in the 16 realizations of numerical simulations using Intervening Opportunities (c) and Gravity (d) models of migration. 
Symbols denote the median number of clusters at a given time step and error bars correspond to the 10th and 90th percentile. 
While for the Gravity model (d)
the number of clusters at stationarity is independent of $n_0$, for the IO model (c) the number of clusters becomes larger as $n_0$ and the total population increase.
{\bf e},  Zeroth law of urbanisation. According to our deterministic model of population dynamics, Eq.\ref{eq:dyn}, cities can form if the population is sufficiently mobile, i.e. if the migration rate $T$ is sufficiently higher than the population growth rate $g$. The three curves show the conditions for the equilibrium state of uniform population to be unstable for the exponential IO model, Gravity models, and a power-law IO model (from bottom to top). 
{\bf f},  The wavelength of the mode with the highest instability, $k_m$, which is proportional to the number of cities per unit length, as a function of the ratio $\rho_0/w$. 
For Gravity models (red curves) $k_m$ tends to a constant in the limit of high carrying capacity, $\rho_0 \gg w$, whereas in the same limit $k_m$ grows proportionally to $\rho_0$ for IO models (green curves). 
} \label{fig:3}}
\end{figure}

Ranking cities by population, it has been observed\cite{RefWorks:239,RefWorks:242} that 
the population of the $i$-th largest city of a country is approximately equal to the population of the largest city divided by $i$, i.e. 
a city's rank is inversely proportional to its population.
In other words, the fraction of cities with population larger than $X$ 
follows a Zipf law, $P_{>}(X) \propto 1/X$.
Several mechanisms have been proposed to account for this observation\cite{RefWorks:229,RefWorks:188,RefWorks:187,RefWorks:151,RefWorks:243}, and the prevailing explanation suggests that the Zipf law results from a stochastic process based on the rule of proportionate random growth, or Gibrat's law\cite{RefWorks:237,RefWorks:339,RefWorks:340}: 
the growth rate of a city's population is independent of its size.  
Although providing a convincing explanation to the Zipf law, the rule of proportionate random growth is unable to answer the following fundamental questions about the urbanisation process: How do cities form? What determines the number of cities in a country? What is the spatial distribution of these cities? 
To answer these questions we analyse  a comprehensive dataset on the population and location of cities globally\cite{RefWorks:295}. 
We first consider the relationship between the number of cities in a geographical region (country or state) and other relevant quantities like the region's area, population, and density. 
Each circle in Fig.~\ref{fig:1}a corresponds to one of the 210 countries of the world, and the number of cities with more than 50,000 inhabitants in each country ($y$ axis) is displayed as a function of its total population ($x$ axis), population density (colour), and total area (circle size). 
The correlation coefficients between the number of cities and total population (0.84), area (0.64), and population density (-0.04) indicate a strong linear dependence between the number of cities in a country and its total population. 
This hypothesis is further confirmed by considering more homogeneous sets of regions, like the United States in Fig.~\ref{fig:1}b and \ref{fig:2}a, as well as European countries and India's states (Supplementary Information).
In general, there is clear evidence that the number of cities grows proportionally with the state or country population, whilst there is a small or indirect relationship between the number of cities and the region's area or population density: %
in the US, the cities-population, cities-area, cities-density correlation coefficients are 0.95, 0.04, and -0.08 respectively. 
Our first law of urbanisation establishes the linear dependence between the number of cities and the population of a region: 
{\em The total number of cities in a region is proportional to the region's population}, $C(N) \sim N$, where $C(N)$ is the number of cities within a region with population $N$. 
Combining this result with Zipf's law we can estimate $C(N, X)$, the number of cities above a given size, $X$, in a region with total population $N$ as 
\begin{align}\label{eq:1st_law}
	C(N, X) = C(N) P_>(X) \sim N/X.
\end{align}
In Figure~\ref{fig:1}c we plot the number of cities with more than $X$ inhabitants in each US state, $C(N,X)$, as a function of the ratio $N/X$ for values of $X$ ranging from 5,000 to 5,000,000 inhabitants. All points collapse on a straight line, confirming that Eq.~\ref{eq:1st_law} holds for several orders of magnitude of $N$ and $X$. 
The first law of urbanisation is also a condition to ensure the stationarity of the Zipf law for a country with a uniformly growing population. Indeed, assuming that the stationary distribution of city sizes follows Zipf's law, if the population of each city doubles in a time $\Delta t$, and so does the country's total population, then the number of cities must also double, otherwise the city size distribution would shift to the right and not be stationary.  
Thus when the total population of a region increases, not only do existing cities increase in size but new cities are also created in accordance with Eq.~\ref{eq:1st_law}.
We find evidence of this behaviour in Iowa, US, where 
historical data %
shows that between 1850 and 2000 the number of incorporated places (i.e. self-governing cities, towns, or villages) grew at the same rate as the state population (Figure~\ref{fig:2}b). 
The second law of urbanisation describes the spatial distribution of cities: 
{\it The distribution of cities in space is a statistically self-similar object with fractal dimension equal to 2}. 
This means that the average number of cities in a region with uniform population density (if measured on a length scale larger than the average distance between cities) is proportional to the region's area, or equivalently that the density of cities scales as $\chi \sim C/A$. 
Combining this result with the first law and Zipf's law, Eq.\ref{eq:1st_law}, and observing that the average distance to the closest city, $\langle d_c \rangle$, scales as the inverse of the square root of the density of cities, we obtain the following result: 
\begin{align}\label{eq:2nd_law}
	\langle d_c \rangle \sim 1 / \sqrt{\chi} \sim \sqrt{A / C} \sim  \sqrt{X (A/ N)} \sim \sqrt{X / \rho} 
\end{align}
i.e. the average distance to the closest city for cities with more than $X$ inhabitants is proportional to the square root of $X$ and inversely proportional to the square root of the region's population density, $\rho \equiv N/A$. 
Figure~\ref{fig:2}c shows the average distance to the closest city with more than $X$=5,000 inhabitants for the United States as a function of the state population density, and confirms the scaling behaviour predicted by Eq.\ref{eq:2nd_law}. 
This finding supports some of the conclusions of the Central Place Theory of human geography\cite{RefWorks:252,RefWorks:253}, whilst disproving others. 
On the one hand, it is true that for a region with uniform population density the larger the cities are, the fewer in number they will be, and the greater the distance, i.e. increasing $X$ in Eq.\ref{eq:2nd_law} results in a greater average distance $\langle d_c \rangle$. 
On the other hand, the average distance between cities of a given size $X$ is not the same for all the states, but depends on the state's population density: cities of a given size are closer in densely populated states than in sparsely populated ones, i.e. for a fixed city size $X$ and state area $A$ the distance between cities decreases as the inverse square root of the state population, $N$. 
To explain the empirical laws of urbanisation, Eqs.\ref{eq:1st_law} and \ref{eq:2nd_law}, we must understand the effect of migrations on cities' demographic dynamics. 
Traditionally, two main mechanisms have been used to estimate and model migrations and other aggregated spatial flows: Gravity models\cite{RefWorks:126,RefWorks:40} and Intervening Opportunities (IO) models\cite{RefWorks:158,RefWorks:159,RefWorks:192,RefWorks:302}. 
In both approaches flows are estimated as the product of two types of variable; one type that depends on an attribute of each individual location (the number of opportunities, usually identified with population), and the other type that depends on a quantity relating a pair of locations (i.e. a distance). 
The difference between the two models pertains to the distance variable considered: the geographical distance in Gravity models, and the number of intervening opportunities in IO models (the number of intervening opportunities between locations $i$ and $j$ is defined as the sum of the opportunities of all locations that are closer to $i$ than $j$ is). 
Both models are able to estimate migration flows with comparable accuracy when fitted to empirical data, %
and currently there is no objective quantitative criterion for selecting one modelling approach over the other in order to infer which between geographic distance or intervening opportunities is the variable that best describes domestic migration flows. 
We characterise the patterns of population distribution generated by each model of migration by running two sets of numerical simulations of the spatiotemporal evolution of a population distributed in the cells of a square grid. 
At every time step the population of each cell can vary due to both natural increase (i.e. births-deaths) modelled with the proportionate random growth mechanism, and migrations of individuals from/to other cells modelled using a Gravity model in one case, and an IO model in the other (see Supplementary Information). 
Imposing a reflecting boundary condition to ensure that the population in each cell cannot decrease below a minimum value $n_0$ and
starting from a random population slightly above $n_0$ in each cell, we observe that 
in both cases the total population increases with $n_0$ (see insets of Fig.~\ref{fig:3}a-b) 
and the population distribution of the 
emerging peaks or clusters of high density, i.e. cities, %
is close to Zipf's law (see main panels of Fig.~\ref{fig:3}a-b). 
The difference between the two migration models is that for the Gravity model the number of clusters is independent of $n_0$, whereas for the IO model, fixing the values of all other parameters, the number of clusters grows with $n_0$ and the total population, see Fig.~\ref{fig:3}c-d. %
To explain this result we consider the corresponding deterministic dynamics, using  an approach similar to the ones developed in economic geography\cite{RefWorks:225,RefWorks:220,RefWorks:298,RefWorks:310}. 
In this approach we represent the spatial distribution of population as a continuous density, and describe its time evolution using a deterministic equation that comprises of two terms: migrations and natural increase.
We model migrations using a continuum version of Gravity and IO models\cite{RefWorks:233} in which the deterrence function, $f$, describing the decay of flows with distance, can be any continuous function that may depend on one or more parameters denoted by $R$. 
We estimate the overall opportunities at a location as the sum of natural resources, $w$, such as fertile soil and minerals which we assume to be constant and uniformly distributed so that unpopulated locations are a priori equivalent, and population-driven opportunities such as jobs and services which are proportional to the population, $\rho$. %
We model natural increase using a logistic equation characterised by two parameters; the carrying capacity, $\rho_0$, which is equal to the stationary density of population attainable in each location in isolation, and the growth rate, $g$, which determines the speed to reach the stationary state. 
The resulting dynamic equation for the density of population $\rho(x,t)$ is 
\begin{align}\label{eq:dyn}
 \dot{\rho}(x,t) = 
    g \rho(x,t) [1 - \rho(x,t)/\rho_0] 
  - T \rho(x,t) 
  + T^{\text{in}}(\rho(t), w, f, R)
\end{align}
where the constant $T$ is the migration rate, and the term $T^{\text{in}}$ represents the increase in density at location $x$ due to migrations from all other locations (see Supplementary Information). 
We observe that a uniform density of population $\rho(x)=\rho_0$ is an equilibrium state because the growth term vanishes and the net migration at every location $T^{\text{in}} - T \rho_0$ is zero by symmetry. 
First, we determine the conditions for the formation of cities by linearising Eq.\ref{eq:dyn} around $\rho_0$ and studying its stability. 
If $\rho_0$ is an unstable equilibrium then an initially small perturbation can grow leading to the formation of zones of high population density (cities), whereas if the state of uniform density $\rho_0$ is a stable equilibrium no cities will form. 
For Gravity models in one and two dimensions, the uniform state is unstable and cities emerge when $T > 4 g \frac{\rho_0 (\rho_0 + w)}{(\rho_0 - w)^2}$, which tends to $T > 4 g$ in the limit of large carrying capacity $\rho_0 \gg w$ (Fig.~\ref{fig:3}e). %
This means that cities will form if the population is sufficiently mobile, i.e. if the migration rate $T$ is sufficiently higher than the growth rate $g$, and this result is independent of both the particular deterrence function $f$ considered, and the average distance of migrations, determined by $R$.
For IO models in one and two dimensions, the condition for the instability does depend on the particular deterrence function considered (see Supplementary Information), however in the large population limit $\rho_0 \gg w$ we recover the same result as the Gravity model: cities form if the population is sufficiently mobile, $T > B g$ where constant $B$ may depend on $f$ (Fig.~\ref{fig:3}e).
We can thus conjecture a zeroth law of urbanisation: 
{\it The formation of cities is only possible if the migration rate is sufficiently higher than the rate of natural increase.}
Second, we determine the number of cities that will form from an unstable equilibrium by computing the wavelength of the mode with the highest instability, $k_m$, which is proportional to the number of cities per unit length (i.e. inversely proportional to the characteristic distance between cities). 
The two migration models produce radically different results. %
For Gravity models the number of cities, $k_m$, depends on the range of migrations determined by the deterrence function $f$ and its parameters $R$, but is independent of the carrying capacity in the limit $\rho_0 \gg w$; 
for IO models $k_m$ depends on $f$, $R$, and grows proportionally to the total population $\rho_0$ in the limit $\rho_0 \gg w$ (see Fig.\ref{fig:3}f and Supplementary Information), 
in agreement with numerical simulations and the laws of urbanisation.  
The fact that Gravity and IO models generate different spatial patterns of population distribution enables the possibility to assess their compatibility with the empirical laws of urbanisation, hence providing a criterion to determine which between geographic distance or number of intervening opportunities is the correct variable to describe migration flows. 

\bibliographystyle{naturemag}

\newpage

\newpage

\newpage

\newpage


\section*{Supplementary material}

# Discovering the Laws of Urbanisation - Supplementary Information

# Contents

# 1 The Dynamic Equation

A general theory aiming at describing human demographic dynamics, i.e. the spatio-temporal evolution of the population distribution in a region, must consider all factors that contribute to the population change in any location. These contributions are reduced to the following two fundamental categories: births, deaths and external (international) migrations on one end, and internal (national) migrations or relocations on the other. The first contribution modifies the total population of the region, while the second is responsible for the redistribution of individuals within the region's area.

If relocations from one point to another are to be considered, alongside population growth and international migrations, a deterministic equation to describe the change in population density, $\rho(x,t)$, at location $x$ is given by:

$$\frac{\partial \rho(x,t)}{\partial t} = g(\rho(x,t)) - T^{out}(x,t,\rho) + T^{in}(x,t,\rho). \tag{S1}$$

$T^{out}$ and $T^{in}$ are the number of individuals leaving and relocating to point $x$ from other parts of the region respectively; these terms represent internal migrations. The form of these functions is determined by the mathematical model used to describe the flow of individuals who relocate from one point to another. Here we consider a Gravity model alongside Intervening Opportunities and Radiation models. The probability that an individual will relocate to a given location is dependent on both the distance to and opportunities at the destination (Gravity models) or at all intermediate locations (Intervening Opportunities and Radiation models). Here, an opportunity refers to any property of a location that may be of interest to the relocating individual. We distinguish between two classes of opportunities; those that depend on the population, such as availability of goods and services, and those that are independent of population such as the presence of naturally occurring resources.

The function $g(\rho(x,t))$ represents the contribution in population change caused by births, deaths, and external (international) migrations. It can be zero, $g(\rho(x,t)) = 0$, if the system has constant population, linear $g(\rho(x,t)) = g \cdot \rho(x,t)$ for an exponential population growth, or Logistic $g(\rho(x,t)) = g \cdot \rho(x,t)(1 - \rho(x,t)/\rho_0)$, characterizing population dynamics with a uniform stationary state, $\rho(x,t) = \rho_0$.

In the following we consider a logistic form of growth and study the time evolution of a continuous population distribution when subjected to small perturbations about the uniform stationary state $\rho(x,t) = \rho_0$. While this approach neglects the inherent randomness of stochastic models required to reproduce phenomena such as Zipf's law, it allows for a fully deterministic analysis of the migration models and their parameters, particularly the parameter constraints under which cities will form. Furthermore we note that additional contributions, such as a diffusion term responsible for the relocation of individuals from highly populated areas to less populated suburbs, can easily be added, however these will not be considered here.

# 2 Gravity Model

Gravity models are derived from Newton's Law of Gravitation;

> *Any two bodies attract one another with a force that is proportional to the product of their masses and inversely proportional to the square of the distance between them.*

In the context of migratory dynamics, the Gravity model suggests that the probability per unit time that an individual moves between two locations, say $i$ and $j$, is proportional to the product of the population densities at $i$ and $j$, each to some power, and a function of the distance, $r_{ij}$, between $i$ and $j$, ie:

$$P_i(j) = \frac{\rho(i)^\alpha \rho(j)^\beta f(r_{ij})}{\int_D dj \rho(i)^\alpha \rho(j)^\beta f(r_{ij})}. \tag{S2}$$

The denominator ensures that the probability, $P_i(j)$, is correctly normalised over the domain D. Exponents $\alpha$ and $\beta$ are usually positive, and often $\beta = 1$. $f(r)$ may be any continuous function, usually an exponential, $e^{-rR}$, or power law, $r^{-\gamma}$.

In Equation S2 it is assumed that individuals will relocate to another position only if there is non-zero population at the destination. In reality, the probability for an individual to relocate is also dependent upon the availability of natural resources at the new location or more generally, any opportunity which is not related to the population, such as the presence of fertile soil or minerals. If these resources are combined into a single term, $w$, then we can re-write Equation S2 as

$$P_i(j) = \frac{[\rho(j) + w(j)] f(r_{ij})}{\int_D dj [\rho(j) + w(j)] f(r_{ij})}. \tag{S3}$$

where $\beta$ has been replaced with 1 and $\rho(j) + w(j)$ denotes the total number of opportunities at location $j$. For simplicity we assume that $w$ is stationary in time; this may be justified by the fact that the rate of recovery of such resources is greater than the rate of consumption.

In order to model the change in population density at location $i$, ie $\rho(i)$, in time, the average number of people both leaving from and moving to $i$ must be included. The number of people leaving $i$ per unit time is given by:

$$T^{out}(i) = T(\rho(i), w(i))\rho(i) \int_D dj P_i(j) = T(\rho(i), w(i))\rho(i). \tag{S4}$$

In the above, $T(\rho(i), w(i))$ is the average migration rate, i.e. the fraction of people in $i$ that will relocate in a time unit. To a first approximation this is assumed to be constant and independent of location; $T(\rho(i), w(i)) = T$. The average number of people relocating to $i$ per unit time is given by:

$$\begin{aligned}
T^{in}(i) &= \int_D dj T(\rho(j), w(j))\rho(j) P_j(i) \\
&= [\rho(i) + w(i)] \int_D dj \frac{T(\rho(j), w(j))\rho(j) f(r_{ij})}{\int_D di [\rho(i) + w(i)] f(r_{ij})}.
\end{aligned} \tag{S5}$$

Inserting these expressions into Equation S1 allows for the change in population density with time, as described by a Gravity model, to be studied.

## 2.1 Pattern Formation and Growth in 1D

Considering a 1-dimensional line of infinite length, we write the dynamic equation, S1, for the Gravity model as:

$$
\begin{aligned}
\frac{\partial \rho(x,t)}{\partial t} =& g(1 - \frac{\rho(x,t)}{\rho_0})\rho(x,t) - T\rho(x,t) \\
&+ T(\rho(x,t) + w)\Big(\int_0^\infty \frac{f(r)\rho(x-r,t)}{\int_0^\infty f(z)(\rho(-r+x-z,t) + \rho(-r+x+z,t) + 2w)\,dz} \\
&+ \frac{f(r)\rho(r+x,t)}{\int_0^\infty f(z)(\rho(r+x-z,t) + \rho(r+x+z,t) + 2w)\,dz}\,dr\Big).
\end{aligned}
\tag{S6}
$$

The uniform distribution $\rho(x) = \rho_0$ is a stationary state of Equation S1 because the growth term is equal to zero and $T^{in}$ is equal to $T^{out}$ for all $x$, hence the time derivative on the left-hand side is zero. In order to determine the stability of the uniform steady state we linearise Equation S6 about $\rho_0$ and study the time evolution of a small perturbation $\tilde{\rho}(x,t)$;

$$
\begin{aligned}
\frac{\partial \tilde{\rho}(x,t)}{\partial t} =& (-g - T)\tilde{\rho}(x,t) + T\int_0^\infty \frac{2\rho_0 f(r)\tilde{\rho}(x,t)}{\int_0^\infty f(z)(2\rho_0 + 2w)\,dz}\,dr \\
&+ T(u+w)\int_0^\infty \Big(-\frac{\rho_0 f(r)\int_0^\infty f(z)(\tilde{\rho}(-r+x-z,t) + \tilde{\rho}(-r+x+z,t))\,dz}{(\int_0^\infty f(z)(2\rho_0 + 2w)\,dz)^2} \\
&- \frac{uf(r)\int_0^\infty f(z)(\tilde{\rho}(r+x-z,t) + \tilde{\rho}(r+x+z,t))\,dz}{(\int_0^\infty f(z)(2\rho_0 + 2w)\,dz)^2} \\
&+ \frac{f(r)\tilde{\rho}(x-r,t)}{\int_0^\infty f(z)(2\rho_0 + 2w)\,dz} + \frac{f(r)\tilde{\rho}(r+x,t)}{\int_0^\infty f(z)(2\rho_0 + 2w)\,dz}\Big)\,dr.
\end{aligned}
\tag{S7}
$$

$\tilde{\rho}(x,t)$ represents a small fluctuation in the stationary density, ie $\rho(x,t) = \rho_0 + \tilde{\rho}(x,t)$, and can be decomposed as a sum of normal modes of wave number $k \in \mathbb{R}$ and amplitude $h_k(t)$.

Substituting $\tilde{\rho}(x,t) = h_k(t)e^{ikx}$ into Equation S7 we obtain:

$$
\begin{aligned}
\frac{\partial h_k(t)}{\partial t} =& (-g - T)h_k(t) + \frac{T\rho_0 h_k(t)\left(\int_0^\infty f(r)\,dr\right)}{\int_0^\infty f(z)(\rho_0 + w)\,dz} \\
&+ T(\rho_0 + w)h_k(t)\left(\frac{\int_0^\infty f(r)\cos(kr)\,dr}{\int_0^\infty f(z)(\rho_0 + w)\,dz} - \frac{\rho_0\left(\int_0^\infty f(r)\cos(kr)\,dr\right)^2}{(\int_0^\infty f(z)(\rho_0 + w)\,dz)^2}\right),
\end{aligned}
\tag{S8}
$$

which may be written as

$$
\frac{\partial h_k(t)}{\partial t} = h_k(t)\Lambda_k(\rho_0, g, T, w, f).
\tag{S9}
$$

Here $\Lambda_k$ is the growth rate of the instability of the normal mode with wavenumber $k$; in order for cities to develop and patterns to emerge $\Lambda$ must be greater than zero for some $k$. We consider the cases in which the deterrence function, $f(r)$ takes an exponential or power law form.

### 2.1.1 Exponential Deterrence Function

If $f(r)$ takes an exponential form (ie. $f(r) = e^{-rR}$) then the growth function $\Lambda$ is given by:

$$\Lambda_k(\rho_0, g, T, w, R) = -g - T\frac{\rho_0}{\rho_0 + w} + T\frac{w + (k/R)^2(\rho_0 + w)}{(1 + (k/R)^2)^2(\rho_0 + w)}. \tag{S10}$$

This expression simplifies using rescaled variables to give:

$$\lambda_\kappa(\mu_0, \tau) = -1 - \tau\frac{\mu_0}{\mu_0 + 1} + \tau\frac{1 + \kappa^2[\mu_0 + 1]}{[1 + \kappa^2]^2[\mu_0 + 1]}. \tag{S11}$$

Here, $\kappa = k/R$, $\tau = T/g$, $\mu_0 = \rho_0/w$ and $\lambda = \Lambda/g$.

The curve defined by Equation S11 is displayed in Figures 1a and 1b for different values of $\mu_0$ and $\tau$.

As aforementioned, in order for cities to form, $\lambda_\kappa$ must be greater than zero at its maximum $\kappa$. The maximum is found by differentiating the expression for $\lambda_\kappa$ with respect to $\kappa$. On doing this, we find that $\lambda_\kappa(\mu_0, \tau)$ has a maximum in $\kappa_m=0$ of $\mu_0 \leq 1$ and in

$$\kappa_m = \sqrt{(\mu_0 - 1)/(\mu_0 + 1)} \tag{S12}$$

otherwise. This corresponds to a wavenumber $k_m = R\sqrt{(\rho_0 - w)/(\rho_0 + w)}$ which, providing $\Lambda_{k_m} > 0$, corresponds to the average number of cities per unit length. For the case that $\rho_0 \gg w$ it can be seen that $k_m$ only depends on $R$, therefore the density of cities is fully determined by the characteristic length of travels, $R^{-1}$, and is independent of all other variables such as growth and migration rate.

The condition on the migration parameter $T$ in order for $\lambda_\kappa$ to be greater than zero is found by substituting the expression for $\kappa_m$ back into Equation S11 and solving for $\tau$. On doing this, the conditions:

$$T_c \geq 4g\rho_0\frac{(\rho_0 + w)}{(\rho_0 - w)^2}, \quad \tau_c \geq 4\mu_0\frac{(\mu_0 + 1)}{(\mu_0 - 1)^2} \tag{S13}$$

are obtained. From these relations we can conclude that if $\rho_0 \gg w$ then cities will emerge if $\tau > 4$, or $T > 4g$. It is important to note that Equation S13 is independent of R; the characteristic travel distance has no effect on whether or not cities will emerge.

### 2.1.2 Power Law Deterrence Function

The power law form for the deterrence function is given by $f(r) = (1 + r)^{-\gamma}$ with $\gamma > 0$. The analysis of Equation S8 is more complicated for this case, however the maximum of $\lambda_k(\mu_0, \tau, \gamma)$ will occur at a $\kappa_m$ that is only dependent on $\mu_0$ and the function $f$ with its parameter $\gamma$.

This is a general result for every function $f$ due to the fact that $\kappa_m$ is defined as $\lambda'_{k_m}(\mu_0, \tau, \gamma) = 0$, with

$$\begin{aligned}\lambda'_k &= \frac{d}{dk}\left\{-1 - \tau\frac{1}{\mu_0 + 1} + \tau\left[\frac{\int_0^\infty f(r)cos(kr)dr}{\int_0^\infty f(z)dz} - \frac{\mu_0}{\mu_0 + 1}\left(\frac{\int_0^\infty f(r)cos(kr)dr}{\int_0^\infty f(z)dz}\right)^2\right]\right\} \\ &= \tau\left[\frac{d}{dk}\left(\frac{\int_0^\infty f(r)cos(kr)dr}{\int_0^\infty f(z)dz}\right) - \frac{\mu_0}{\mu_0 + 1}\frac{d}{dk}\left(\frac{\int_0^\infty f(r)cos(kr)dr}{\int_0^\infty f(z)dz}\right)^2\right],\end{aligned} \quad (S14)$$

a function of $\mu_0$ and $f$ only. For the case of $\mu_0 \gg 1$ (that is $\rho_0 \gg w$), it is only the function $f$, defining the spatial range of the migration process, that the characteristic distance between cities is dependent upon.

$\lambda_{k_m} = 0$ may be solved numerically for different values of $\gamma$. On doing this we find that the critical curve collapses on the same universal function, independent of $\gamma$.

Moreover, it turns out that the critical curve of Equation S13 determining the condition for the formation of cities is valid for any continuous function $f$, including a power law deterrence function and therefore holds universally. We may prove this by defining $A(k) \equiv \left(\int_0^\infty f(r)cos(kr)dr\right)/\left(\int_0^\infty f(z)dz\right)$. From Equation S14 we have

$$\begin{aligned}0 &= A'(k_m) - \frac{2\mu_0}{\mu_0 + 1}A(k_m)A'(k_m) \\ A(k_m) &= \frac{\mu_0 + 1}{2\mu_0}.\end{aligned} \quad (S15)$$

Inserting this into $\lambda_{k_m}(\mu_0, \tau_c) = 0$, the critical curve defined in Equation S13 is re-obtained.

### 2.1.3 1D Gravity - Summary

If a population is described by a logistic growth with individuals relocating according to a Gravity model, cities will develop if the following two conditions are met simultaneously:

- The average population is sufficiently large: $\rho_0 \gg w$
- The population is sufficiently mobile: $T > 4g$

When the population density is sufficiently high, the average distance between cities depends only on the deterrence function $f$ and is independent of growth rate, migration rate and average population.

## 2.2 Pattern formation and Growth in 2D

In 2 dimensions, Equation S1 describing the evolution of the density of population for a Gravity model of human migration is given by:

$$\frac{\partial \rho(x,y,t)}{\partial t} = g\rho(x,y,t)\left(1 - \frac{\rho(x,y,t)}{\rho_0}\right) - T\rho(x,y,t) + T(\rho(x,y,t) + w).$$
$$\left(\int_0^\infty \int_0^{2\pi} \frac{f(r)\rho(r\cos(\theta)+x, r\sin(\theta)+y, t)}{\int_0^\infty \int_0^{2\pi} f(z)(\rho(r\cos(\theta)+x+z\cos(\phi), r\sin(\theta)+y+z\sin(\phi), t)+w)d\phi dz} d\theta dr\right).$$
(S16)

We linearize this equation about the stationary solution, $\rho_0$, obtaining an expression for the evolution of a small perturbation in the population density:

$$\frac{\partial h_{k,l}(t)}{\partial t} = h_{k,l}(t)\left\{T(\rho_0+w)\int_0^\infty \int_0^{2\pi} \left(\frac{f(r)e^{i(kr\cos(\theta)+lr\sin(\theta))}}{\int_0^\infty 2\pi f(z)(\rho_0+w)\,dz}\right.\right.$$
$$\left.-\frac{\rho_0 f(r)\int_0^\infty 2\pi f(z)J_0\left(\sqrt{k^2+l^2}z\right)e^{i(kr\cos(\theta)+lr\sin(\theta))}\,dz}{\left(\int_0^\infty 2\pi f(z)(\rho_0+w)\,dz\right)^2}\right)d\theta dr \quad (S17)$$
$$\left.+T\rho_0\int_0^\infty \frac{f(r)}{\int_0^\infty f(z)(\rho_0+w)\,dz}\,dr - g - T\right\}$$

where $h_{k,l}$ is the amplitude of a 2 D plane wave with wavevector components $k,l$ and $J_0$ is a Bessel function of the first kind.

### 2.2.1 Exponential Deterrence Function

For the 2D case, the exponential form of $f(r)$ is given by $f(r) = e^{-Rr}$, where $r = \sqrt{x^2+y^2}$, the classic Euclidean distance. Substituting this into S17, we obtain an expression for the growth function $\Lambda_{k,l}$:

$$\Lambda_{k,l} = T\left(\frac{\rho_0\left(k^2+l^2\right)}{(\rho_0+w)\left(k^2+l^2+R^2\right)} + \frac{R}{\sqrt{k^2+l^2+R^2}} - 1\right) - g. \quad (S18)$$

The function $\Lambda_{k,l}$ depends only on the magnitude of the wavevector, $p = \sqrt{k^2+l^2}$, and substituting $p^2$ for $k^2+l^2$ in Equation S18 results in an identical expression as that for $\Lambda_k$, Equation S10. Hence the critical value of the ratio between migration and growth above which cities are able to form is the same for both the 1D and 2D Gravity models, given by Equation S13.

### 2.2.2 Power Law Deterrence Function

The power law form of the deterrence function in 2D is given by $f(r) = (1+r)^{-\gamma}$, as for the 1D case, but with $r$ equal to the Euclidean distance. Substituting this into Equation S17 we find that an analytical expression for $\Lambda$ is not obtainable. Solving the resulting equation for the formation

of cities numerically, in analogy with the 2D model with an exponential $f(r)$, the critical curve of Equation S13 is obtained. Summarising, in both 1 and 2 dimensions, if the population dynamics are described by a logistic growth with individuals relocating according to a Gravity model cities will form if $T > 4g$ and $\rho_0 \gg w$ and this is independent of the deterrence function considered.

## 3   Intervening Opportunities & Radiation Models

The Intervening Opportunities model of human migration was hypothesised by Stouffer [1]. His law states:

> The number of persons going a given distance is directly proportional to the number of opportunities at that distance and inversely proportional to the number of intervening opportunities.

According to this model, the probability per unit time of observing an individual at location $i$ moving to location $j$ is given by:

$$P_i(j) = [\rho(j) + w(j)]f\left(\int_{B_i(r_{ij})} dz[\rho(z) + w(z)]\right), \tag{S19}$$

where $B_i(r_{ij})$ is a ball of radius $|j-i|$ centred on $i$ and $f$ is a continuous function. The normalisation of Equation S19 over a domain $D$ with total population $N$ may be imposed as a condition on $f$:

$$\begin{aligned}
\int_D dj P_i(j) &= \int_0^\infty dr\, r \int_0^{2\pi} d\theta[\rho(r,\theta) + w(r,\theta)]f\left(\int_0^r dz\, z \int_0^{2\pi} d\theta[\rho(z,\theta) + w(z,\theta)]\right) \\
&= \int_0^\infty dr \frac{da(r)}{dr} f(a(r)) \\
&= \int_0^N da\, f(a) \\
&= F(0) - F(N) = 1.
\end{aligned} \tag{S20}$$

Here, $a(r) = \int_0^r dz\, z \int_0^{2\pi} d\theta[\rho(z,\theta) + w(z,\theta)]$ and $F(a) \equiv \int_a^\infty du\, f(u)$ is a decreasing function such that $F(0) = 1$ and $F(N) = 0$. The deterrence function, $F(a)$ may be either exponential, $e^{-aR}$, as is the case for the Intervening Opportunities model, or power law ($\frac{1}{1+a}$) which corresponds to the Radiation model.

For these models, the total number of travellers leaving from $i$ per unit time is the same as for the Gravity model, described by Equation S4. The total number of people moving to $i$ per unit time is given by:

$$\begin{aligned}
T^{in}(i) &= \int_D dj\, T(\rho(j), w(j))\rho(j)P_j(i) \\
&= [\rho(i) + w(i)] \int_D dj\, T(\rho(j), w(j))\rho(j)]f\left(\int_{B_i(r_{ij})} dz[\rho(z) + w(z)]\right).
\end{aligned} \tag{S21}$$

These terms combine to give the general dynamic equation for the time evolution of the population density as per Equation S1.

## 3.1 Pattern formation and Growth in 1D

The full 1D dynamic equation for the Intervening Opportunities and Radiation models is given by

$$\begin{aligned}\frac{\partial \rho(x,t)}{\partial t} =& g\rho(x,t)[1 - \frac{\rho(x,t)}{\rho_0}] - T\rho(x,t) + T[\rho(x,t) + w].\\ & \int_0^\infty \left[\rho(x-r,t)f\left(\int_{x-2r}^x [\rho(z,t)+w]dz\right) + \rho(x+r,t)f\left(\int_x^{x+2r}[\rho(z,t)+w]dz\right)\right]dr.\end{aligned} \quad (S22)$$

Using the same method as used for the Gravity model, we write this equation for small fluctuations in $\rho(x,t)$, denoted $\tilde{\rho}(x,t)$, by linearising about the uniform stationary state $\rho_0$:

$$\begin{aligned}\frac{\partial \tilde{\rho}(x,t)}{\partial t} =& -(g+T)\tilde{\rho}(x,t) + 2T\rho_0 \tilde{\rho}(x,t) \int_0^\infty f(2r[\rho_0+w])dr + \\ & T[\rho_0+w]\int_0^\infty \bigg( f(2r[\rho_0+w])[\tilde{\rho}(x-r,t)+\tilde{\rho}(x+r,t)] + \\ & \rho_0 f'(2r[\rho_0+w])\int_{x-2r}^{x+2r}\tilde{\rho}(z,t)dz\bigg)dr.\end{aligned} \quad (S23)$$

Studying the stability of this equation to a small perturbation with specific wavenumber $k$, ie $\tilde{\rho}(x,t) = h_k(t)e^{ikx}$, we obtain a general expression for the growth function, $\Lambda$.

$$\begin{aligned}\Lambda_k =& -(g+T) + \frac{T\rho_0}{\rho_0+w} + \\ & 2T[\rho_0+w]\left[\int_0^\infty \cos(kr)f(2r[\rho_0+w])dr + \rho_0 \int_0^\infty \frac{\sin(2kr)f'(2r[\rho_0+w])}{k}dr\right].\end{aligned} \quad (S24)$$

Introducing the rescaled variables: $\tau = T/g$, $\mu_0 = \rho_0/w$, $\kappa = k/w$ and $\lambda = \Lambda/g$, the above expression becomes:

$$\begin{aligned}\lambda_\kappa(\mu_0,\tau,f) =& -(1+\tau) + \frac{\mu_0}{\mu_0+1} + 2\tau[\mu_0+1].\\ & \left[\int_0^\infty \cos(\kappa z)f(2z[\mu_0+1])dz + \mu_0\int_0^\infty \frac{\sin(2\kappa z)f'(2z[\mu_0+1])}{\kappa}dz\right].\end{aligned} \quad (S25)$$

### 3.1.1 Exponential f(a) - Intervening Opportunities Model

For the Intervening Opportunities model, the deterrence function takes the form of an exponential; $f(a) = Re^{-Ra}$. With this, Equation S25 is evaluated exactly to give

$$\lambda_{\tilde{\kappa}}(\mu_0,\tau) = -(1+\tau) + \tau\frac{\mu_0}{\mu_0+1} - \tau(\mu_0+1)\left[\frac{\mu_0}{\tilde{\kappa}^2+(\mu_0+1)^2} - \frac{4(\mu_0+1)}{\tilde{\kappa}^2+4(\mu_0+1)^2}\right], \quad (S26)$$

where $\tilde{\kappa} = \kappa/R = k/(wR)$. The curve described by Equation S26 is shown in Figures 1d and 1e for different values of the parameters.

The function $\lambda_{\tilde{\kappa}}(\mu_0, \tau)$ has a maximum in $\tilde{\kappa}_m = 0$ if $\mu_0 \leq 1/3$ or in

$$\tilde{\kappa}_m = \sqrt{\frac{-4\mu_0(\mu_0 + 2) + 6\sqrt{\mu_0(\mu_0 + 1)^5} - 4}{3\mu_0 + 4}} \tag{S27}$$

otherwise. The parameter $\tilde{\kappa}_m$ corresponds to a wavenumber $k_m = Rw\tilde{\kappa}_m$. If $\lambda_{k_m} > 0$, cities are able to form and $k_m$ is proportional to the number of cities per unit length. From the expression for $k_m$ we find that, in contrast to the Gravity model, as $\mu_0 \to \infty$ (ie $\rho_0 \gg w$), $k_m \to \infty$. This implies that for a fixed $R$, the density of cities will be greater in more populated countries.

The condition for $\lambda_{\tilde{\kappa}_m} > 0$ is found by inserting the expression for $\tilde{\kappa}_m$ into Equation S25 and solving for $\tau$, to obtain an expression for $\tau_c$; if $\tau > \tau_c$, cities will emerge. $\tau_c$ is given by:

$$\tau_c = \frac{3(5\mu_0^3 + 11\mu_0^2 + 7\mu_0 + 4\sqrt{\mu_0(\mu_0 + 1)^5} + 1)}{(\mu_0 + 1)(3\mu_0 - 1)^2}. \tag{S28}$$

It should be noted that this equation is independent of $R$. Furthermore, if $\mu_0 \gg 1$ ($\rho_0 \gg w$), the condition for growth reduces to $\tau > 3$, or $T > 3g$. Here we may draw a parallel with the Gravity model; in both cases cities can only emerge if the migration rate is sufficiently higher than the growth rate.

### 3.1.2 Power Law f(a) - Radiation Model

For the Radiation model, $f(a) = 1/(1 + a)^2$ and, as with a power law deterrence function in the Gravity model, the evaluation of Equation S25 is more complicated. By solving $\lambda'_{\tilde{\kappa}}(\mu_0, \tau, f) = 0$ numerically, we find that $\tilde{\kappa}_m = 0$ if $\mu_0 \leq 1$, and $\tilde{\kappa}_m > 0$ otherwise. $\lambda_{\tilde{\kappa}_m}(\mu_0, \tau) = 0$ is solved numerically to obtain critical values of the parameters that will allow the emergence of cities.

The results are shown in Figures 2c and 2d where it is seen that the critical curves for the Intervening Opportunities and Radiation models are not the same; in contrast to Gravity models, for Intervening Opportunities type models these curves do depend on the deterrence function used.

A distinct contrast between the Gravity model and an Intervening Opportunities model is the variation of $k_m$ with increasing density $\rho_0$. The value of $k_m$ may be related to the number of cities, $C$, that will grow from a small perturbation through the equation

$$C = Lk_m/2\pi. \tag{S29}$$

For the Gravity model, as $\mu_0 \to \infty$, $k_m$ approaches unity, whereas for the Intervening Opportunities model, $k_m$ increases proportionally to $\mu_0$. This result is displayed in Figure 3f (main text). From this difference is it possible to determine which model of migration best describes patterns seen in data.

### 3.1.3 1D IO and Radiation - Summary

If a population grows following a logistic equation and relocates according to an Intervening Opportunities or Radiation model, an initial perturbation to the steady state population distribution will result in the development of urban settlements if the following two conditions are met simultaneously:

- The average population is sufficiently large: $\rho_0 > C_0 w$, with $C_0 = 1/3$ and 1 for the Intervening Opportunities and Radiation models respectively.
- The population is sufficiently mobile: $T > C_1 g$ with $C_1 = 3$ and 5 for the Intervening Opportunities and Radiation models respectively.

## 3.2 Pattern formation and Growth in 2D

The Intervening Opportunities model in 2 dimensions is studied by considering the equation:

$$\frac{\partial \rho(x,y,t)}{\partial t} = g\rho(x,y,t)\left(1 - \frac{\rho(x,y,t)}{u}\right) - T\rho(x,y,t) + T(\rho(x,y,t)+w). \tag{S30}$$

$$\left(\int_0^\infty \int_0^{2\pi} \rho(x+r\cos(\theta), y+r\sin(\theta), t)\right.$$

$$\left. f\left[\int_0^r \int_0^{2\pi} (\rho(x+r\cos(\theta)+z\cos(\phi), y+r\sin(\theta)+z\sin(\phi), t)+w)d\phi dz\right] d\theta dr\right).$$

This is analogous to the equation in 1 dimension, Equation S22, however now we consider the opportunities within a circle of radius $r$, with $r$ being the Euclidean distance between the origin and destination.

Performing linear stability analysis on the above equation, with a perturbation to the stationary distribution of the form $e^{i(kx+ly)}$, we find that the conditions for growth of cities are unchanged from the 1-dimensional case if the 1 dimensional wavenumber $k$ is replaced by the magnitude of the 2 dimensional wavevector $p$.

This result is demonstrated in Figures 3a and 3b where all parameters are equal to those used to generate Figures 1d and 1e respectively, for the 1D case. Here, $\tilde{p} = p/R$.

# 4 Numerical Simulations

We simulate both the Gravity and Intervening Opportunities models using numerical methods in order to assess the accuracy of the analytical results derived above.

## 4.1 Gravity Model

To simulate the Gravity model in 1-dimension, we evolve the dynamic Equation S6 in time starting with an initial distribution at $t = 0$ of small random perturbations about the steady state solution $\rho_0$. The integrals are evaluated using the trapezoidal rule. By varying the model parameters, the validity of Equation S13 describing the condition for city formation and the relationship between the wavenumber and the number of cities, Equation S29, is determined. Unless otherwise stated, we rescale variables by setting growth rate and resources, $g$ and $w$ respectively, to equal unity; this allows for the model to be specified with 3 variables rather than 5.

On running simulations with fixed parameters $\rho_0 = 80$, $R = 0.5$ and varying the migration parameter $T$ between 3 and 5, we find that Equation S13 holds; we can predict whether an initial perturbation

11

will grow or decay based on the ratio between the migration and growth rate. We can also simulate the model with fixed parameters $T = 10$ and $R = 1.5$ and increasing carrying capacity $10 \leq \rho_0 \leq 90$. From this we find that increasing the carrying capacity increases the size of cities however the number of cities remains constant, determined by Equation S29. The results of this simulation are displayed in Figure 4.

## 4.2 Intervening Opportunities Model

In order to numerically simulate the Intervening Opportunities model, we make an adjustment to the expression for the probability of observing an individual at $i$ moving to $j$, originally formulated in Equation S19. Rather than using the probability density function $f(a)$, we use the cumulative distribution function, $F(a)$. On doing this, we write the dynamic equation, analogous to Equation S22, as:

$$\begin{aligned}\frac{\partial \rho(x_i)}{\partial t} &= g\rho(x_i)(1 - \frac{\rho(x_i)}{\rho_0}) - T\rho(x_i) + T\sum_j P_{ij} \\ &= g\rho(x_i)(1 - \frac{\rho(x_i)}{\rho_0}) - T\rho(x_i) + T.\sum_j \left[\frac{F(a_{ij}) - F(a_{ij} + \rho_j + w_j)}{F(\rho_i + w_i) - F(N)}\right] \cdot \frac{\rho_j + w_j}{\sum_{k \in R_{ij}} \rho_k + w_k}\end{aligned} \quad (S31)$$

where $R_{ij}$ is the set of locations on the ring of radius $r_{ij}$, centered on $i$, therefore the summation:

$$\sum_{k \in R_{ij}} \rho_k + w_k \quad (S32)$$

is a sum over all locations $k$ at the same distance as $j$ from the origin $i$. We define $D_{ij}$ such that it represents the set of locations within the disc of radius $r_{ij}$ centered in $i$ and the opportunities between but not including locations $i$ and $j$ may be written:

$$a_{ij} = \sum_{k \in D_{ij}} \rho_k + w_k. \quad (S33)$$

Using this formulation, the dynamical equation corresponding to a population relocating according to an Intervening Opportunities model is simulated.

We run simulations for fixed parameters $R$ and $\rho_0$ and varying $T$ between 2 and 4; finding that Equation S28 successfully determines whether cities will form from a perturbation about a stationary distribution. Simulations are also carried out for fixed $T$ and $R$ whilst varying $\rho_0$, the results of which are shown in Figure 5. From this we find that the number of cities increases with $\rho_0$ for an Intervening Opportunities model as expected from Equations S27, S29 and Figure 3f (main text).

## 5 Model Comparison

From both the analytical results and those obtained from simulations, we note that the main contrast between the Gravity and Intervening Opportunities models is their behaviour with increasing $\rho_0$,

specifically how the number of cities changes due to a change in the steady state population density. From the analysis of the Gravity model, we find that the number of cities does not vary with $\rho_0$; this is shown in Equation S12 and supported by Figure 4. In comparison to this, for the Intervening Opportunities model, we find that increasing $\rho_0$ also results in an increase in the number of cities. For this model, the expression for $k_m$ suggests a linear relationship between $\rho_0$ and the number of cities, when all other parameters are fixed, and this is supported for low values of $\rho_0$ by Figure 5.

The dynamic equation, S1, used for each model only differs in the migration term; both models follow a logistic growth. As a result, by setting this growth to equal zero the differences between Gravity driven migration and migration driven by an Intervening Opportunities model may be effectively analysed. Setting the growth to zero to study the number of peaks is also justified by the fact that the number of peaks is proportional to $k_m$; for both models this is independent of the value of $g$.

## 5.1 Number of Peaks vs $\rho_0$

### 5.1.1 1D: g=0

Figures 6a and 6c demonstrate the difference between the two models noted above. Here we simulate both models with an identical initial distribution and equal parameters except for $R$, the value of which is specific to the model being used. It is seen that as $\rho_0$ is increased the final distribution from the Gravity model displays peaks of increasing height where as that for the Intervening Opportunities model displays an increasing number of peaks alongside a less significant increase in height. The full time evolution of the initial perturbation is displayed in Figures 7a,c,e (Gravity model) and 7b,d,f (Intervening Opportunities model).

### 5.1.2 1D: g=1

In Figures 6b and 6d we demonstrate the effect of a small growth on the final distributions for the Gravity and Intervening Opportunity models respectively. We observe that the effect of a non-zero growth rate on the final distribution of cities is that all cities evolve to approximately the same size; $\rho_0$. This is a result of the logistic form of growth, $g \cdot (1 - \rho(x,t)/\rho_0)$; as $\rho(x,t)$ approaches $\rho_0$, the growth rate tends to zero, all cities have reached the same size and have equal opportunities, therefore there is also no further migration. The full time evolution of the initial perturbation is displayed in Figures 8a,c,e (Gravity model) and 8b,d,f (Intervening Opportunities model).

### 5.1.3 2D

Figure 9 displays the result from 2D a simulation of an initially random distribution of population with a logistic growth and individuals relocating according to a Gravity model with an exponential deterrence function. The dynamics for this simulation are described by Equation S16. From this we observe that the initial population distribution, randomised about the value $\rho_0 = 25$, converges on a stationary state with 6 cities of varying population. Figure 10 displays the corresponding evolution for individuals relocating according to an Intervening Opportunities model, described by Equation S30, when the initial distribution is randomised about $\rho_0 = 25$. From this it is evident

that the distribution reaches a steady state with 30 cities of varying size. Both simulations were performed with fixed parameters $L = 15$, $T = 30$, $R = 0.5$ (Gravity model) and $R = 0.2$ (Intervening Opportunities model).

In Figure 11 we display the final stationary distributions of cities obtained from simulations of both the Gravity model (red squares) and Intervening Opportunities model (green triangles) for initial distributions randomised about increasing values of $\rho_0$ over the range [10,50]. From this we observe that, in correspondence with the 1-dimensional simulations, the number of cities for the Gravity model remains fixed and constant with increasing $\rho_0$ where as for the Intervening Opportunities model the number of cities increases with $\rho_0$. As the initial distribution of population is randomised about $\rho_0$ this result demonstrates that if individuals relocate according to a Gravity model the number of cities is unaffected by the total population of the region. In contrast, if it is an Intervening Opportunities model by which people relocate then the number of cities within a region will increase with the total population of that region.

# 6 Assessing the Laws of Urbanisation

Here we present the analysis of data sets for European countries and states within India, with the goal of assessing the validity of the two laws of urbanisation stated in the main text: the relationship between total population and the number of cities within a region (first law, Eq. 1), and the relationship between population density and average distance between cities (second law, Eq. 2).

Figure 12a displays the relationship between the number of cities with over 5,000 inhabitants ($y$ axis) for each individual country within Europe and the country's total population ($x$ axis), population density (colour) and area (size). From this we see that there is strong linear correlation (0.92) between the number of cities with more than 5,000 inhabitants and total population of a country, whilst there is a weak correlation between number of cities and area (0.53) and no correlation between the number of cities and population density (-0.1), thus supporting the first law of urbanisation. Further support for the first law comes from the analysis of data from India; Figure 13a displays the same results as 12a for states within India; the corresponding correlation coefficients between number of cities and total population (0.92), area (0.74), and density (-0.1) confirm the validity of the first law.

In 12c and 13c we present further evidence for the validity of Eq. 1 (main text); for countries or states with a total population $N$, the number of cities with more than $X$ inhabitants, with $X$ ranging from 5,000 to 5,000,000, is a linear function of $N/X$. The implication of this result is that the number of cities of size $X$ within a country or state may be fully determined by the population of the region, $N$.

In Figures 12b and 13b we display the average distance to the closest city with more than $X$=5,000 inhabitants for countries in Europe and more than $X$=1,000 inhabitants for states in India respectively. The strong linear relationship for both data sets provides further support for Eq. 2 (main text); the average distance between cities of population $X$ depends on both $X$ and the population density of the region considered.

From 13b we observe that the Indian state of Goa lies away from the line; 13d shows that this state has a significantly smaller area than other states while 13a displays that, given the total population $N$, there are more cities than expected by the first law of urbanisation. From this we can conclude

that the region has a higher than average number of cities given its population and a smaller than average area, thus explaining why the average distance between cities is lower than expected.

The above analysis is displayed for states within the US in Figure 14 providing further support for the first and second law.

Figure 15 displays the average distance to the closest city with population larger than $X$ rescaled by the square root of $X$ $\langle d_c \rangle X^{-0.5}$ ($y$ axis) vs population density $N/A$ ($x$ axis) for states within the US, countries within Europe and states within India ($X$ ranges from 1,000 to 1,000,000). All points collapse on the line $y = x^{-1/2}$ as prescribed by the second law of urbanisation (Eq. 2 main text) indicating that the average distance between cities with more than $X$ inhabitants scales as the inverse root of the population density (ie. $N/A$) of the region being considered.

In Figure 16 we display the correlations between total population $N$, area $A$, population density $\rho$ and the number of cities with more than $X = 5,000$ inhabitants (black) and $X = 50,000$ inhabitants (red) for states within the US, countries in Europe, states within India. The total population $N$ and the number of cities are the most strongly correlated variables for all regions and both values of $X$, demonstrating the validity of the first law of urbanisation (Eq. 1 main text).

The relatively high correlation between number of cities and area is due to the correlation between the total population of a region and its area; a region with a large area can easily accommodate a large population. The correlations between total population and area for the US, European and Indian regions are displayed in Figure 17.

# 7 Stochastic Simulations

We now consider a stochastic model of population dynamics which combines the models of migration with Gibrat's proportionate random growth. Such an approach is expected to reproduce both the laws of urbanisation and Zipf's law for city sizes.

For this approach, we model the population as individuals rather than a continuous density. Space is discretised into cells of size $(L_x/N_x) \times (L_y/N_y)$; $L_x$ and $L_y$ are the sizes of the 2-dimensional region being considered and $N_x$, $N_y$ are the number of cells in each dimension. For simplicity, we will start by only considering square regions with $L_x = L_y = L$, and $N_x = N_y = N$, thus the total number of cells is $N^2$ and each has an area of $(L/N)^2 \equiv dl^2$. Each cell may be specified by a pair of coordinates $(x, y)$ with $0 \leq x < N$ and $0 \leq y < N$. The population of cell $(x, y)$ at time $t$ is given by $n(x, y, t)$.

The population in any cell may change in two ways; natural increase and migrations. Here, natural increase refers to the difference between the number of births and the number of deaths within a cell therefore it may be both positive or negative. Migrations refers to any person relocating between cells; both inward migrations and outward migrations must be accounted for.

The mechanism that dictates the natural increase in the population of a cell is proportionate random growth, the growth mechanism known to produce Zipf's Law. If we consider the change in the population of cell $(x, y)$ due to births and deaths between time $t$ and $t + dt$, calling this change $\delta_g(x, y, t)$, then by the rule of proportionate random growth we have $\delta_g(x, y, t) = \xi(t) n(x, y, t)$. Here, $\xi(t)$ is a normally distributed random variable with mean $\mu$ and variance $\sigma^2$. In order for the resulting distribution of city sizes to be a power-law [2], $\mu$ must be slightly negative, $\sigma$ must be

small (of order $\approx 0.1$) and there must be a reflecting boundary condition preventing the population of a cell from going below a minimum value, $n_{min} = n_0$.

We model relocations between cells using either a Gravity or an Intervening Opportunities model. To implement this, we define a probability of migration, $T$, which corresponds to the probability that an individual in cell $(x, y)$ will relocate to any other cell between time $t$ and $t+dt$. The total number of outgoing migrants from cell $(x, y)$ in time $dt$, $m(x, y, t)$ is extracted from a binomial distribution with migration probability $T = 0.4$. We then relocate these outgoing migrants according to the Gravity or Intervening Opportunities model as described in the following sections.

## 7.1 Gravity Model

If individuals relocate according to a Gravity model, the probability of a migrant relocating from $(x, y)$ to $(i, j)$ is given by

$$P(x, y \to i, j) = \frac{F(r(x, y \to i, j))[n(i, j, t) + w(i, j, t)]}{\sum_{i,j} F(r(x, y \to i, j))[n(i, j, t) + w(i, j, t)]} \tag{S34}$$

where $r(x, y \to i, j)$ is the euclidean distance between locations $(x, y)$ and $(i, j)$. The deterrence function $F$ is assumed to be exponential, $e^{-rR}$.

## 7.2 IO Model

According to the Intervening Opportunities model, the probability of a migrant relocating from $(x, y)$ to $(i, j)$ is given by

$$P(x, y \to i, j) = \left[ \frac{F(A_{x,y \to i,j}) + F(A_{x,y \to i,j} + A_{i,j})}{F(n(x, y, t) + w(x, y)) - F(N)} \right] \cdot \frac{n(i, j, t) + w(i, j)}{A_{i,j}} \tag{S35}$$

where $A_{i,j}$ corresponds to the sum of the population and opportunities in all cells at an equal distance from $(x, y)$ as $(i, j)$ and F is the deterrence function that we assume to be exponential, $e^{-AR}$. $A_{x,y \to i,j}$ is the sum of the population and resources of all intervening cells; all cells that lie at a distance from $(x, y)$ that is less than the distance to $(i, j)$, including the population and resources of cell $(x, y)$ itself.

Using Equations S34 or S35, we compute the probability of relocation to all cells according to the model in question. The number of migrants moving between any two locations are then extracted from a multinomial distribution. The number of incoming migrants to cell $(x, y)$ in between time $t$ and $t + dt$, $\delta_{in}(x, y, t + dt)$, is given by the sum of the outgoing migrants from all other cells that have destination $(x, y)$. We denote this sum $\Phi(x, y, t)$.

Using this framework, we simulate the population dynamics. In a time $dt$, the population in each cell will change due to natural increase, outgoing migrants and incoming migrants. The population of a cell $(x, y)$ after $dt$ is therefore given by:

$$\begin{aligned} n(x, y, t + dt) &= n(x, y, t) + \delta_g(x, y, d) - \delta_{out}(x, y, d) + \delta_{in}(x, y, d) \\ &= n(x, y, t) + \xi(t)n(x, y, t) - m(x, y, t) + \Phi(x, y, t). \end{aligned} \tag{S36}$$

16

We start our simulation at time $t = 0$ from an initial condition where the population of each cell is drawn from a uniform distribution within the range $n_0 \leq n \leq 1.2 \cdot n_0$, where $n_0$ is the minimum population of a cell.

After each time step, $dt$, the state of the system is assessed by counting the total number of cities and their populations. Cities correspond to clusters of population and may therefore be defined as adjacent populated geographical spaces (cells) [3]. We define a cluster as a set of adjacent cells for which the Manhattan distance of a single cell from the closest cell that is a member of the cluster has a maximum of 1. This satisfies the adjacency condition and is equivalent to a condition on the cell edges; any cell that is a member of a cluster must share at least one edge with another cell within the same cluster. Alongside this, the population of every cell within the cluster must be greater than a minimum value $X$ to ensure that all cells are populated.

Results of simulations with increasing $n_0$ are displayed in Figure 18. Figures 18g and 18h display the time evolution of the number of clusters with $T = 0.4$, $X = 2000$, $\mu = -0.01$, $\sigma = 0.25$, $L = 60$ and $N = 60$ computed according to the above algorithm, for the Intervening Opportunities and Gravity models respectively. From 18g, we observe that, when individuals relocate according to an Intervening Opportunities model the number of clusters (cities) at any given time increases with $n_0$. Conversely, after an initial transient, the number of clusters is independent of $n_0$ when individuals relocate according to a Gravity model (18h).

In Figures 19 and 20 we display the variation in the average total population with simulation time for the Gravity and Intervening Opportunities models respectively. The shaded regions correspond to the values of total population that fall within the 25th and 75th percentiles at each step. For both models we observe that simulations with larger $n_0$ also have a larger total population at all time steps, implying that the total population is an increasing function of $n_0$.

Figures 21 and 22 display the average CCDF of cluster population for increasing $n_0$. The grey line represents the Zipf distribution. Both models of migrations successfully produce clusters with a population distribution fitted by Zipf's Law, however while the Intervening Opportunities model has a distribution that is independent of $n_0$, the Gravity model's distribution shifts to the right as $n_0$ increases.

## 8 Growth

The linear stability analysis (sections 2 and 3) performed on both models of migration is only valid for small perturbations about the stationary distribution. With regards to the simulations, this means that the conditions for the number of cities (Equation S29) are only valid during the initial stages of the simulation; at later times the effects of the non-linearities present in Equation S1 become more pronounced.

From the linear stability analysis it was concluded that the growth and migration rates affect whether or not cities may form from the initial perturbation (Equations S13 and S28) however the number of cities that form depend on the parameters $\rho_0$, $R$ and $w$ only. Numerical simulations indicate that on longer time scales, the number of cities that are present also depends the growth rate $g$ and the migration parameter $T$. In particular the number of cities is constant for a given ratio $T/g = \tau$, however if the ratio is varied, by changing either $g$ or $T$, the number of cities also changes.

We display these results in Figure 23 where it is observed that the number of cities present after a long time period increases with $g$ for both the Gravity model (a-c) and the Intervening Opportunities model (d-f). If the migration parameter $T$ is increased, the number of cities decreases.

In order to find empirical confirmation to the model's prediction, we would need a complete series of historical data on both growth and migration rates. While it is possible to estimate the growth rate of each state within the US since 1790 using census data, we were not able to determine the migration rates of the states prior to 1940, the first year in which the US census included questions about people's mobility [4]. Due to the lack of historical data on migration rates prior to 1940 we are unable to estimate the ratio $T/g$ of each state, hence precluding the possibility to test the presence of a correlation between the ratio $T/g$ and the number of cities.

# 9 Figures

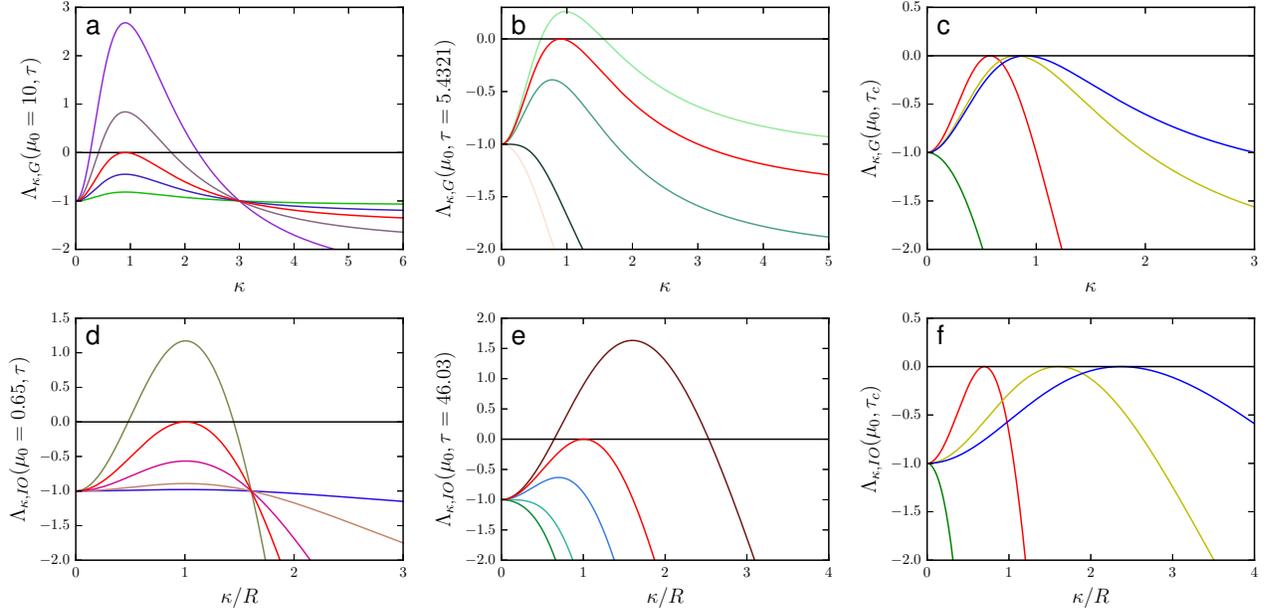

Figure 1: **a** The function $\lambda_\kappa(\mu_0, \tau)$ from Equation S11 with fixed $\mu_0 =10$ and varying $\tau$: 1,3,10,20. The red line corresponds to $\tau_c = 5.4321$, as described by Equation S13. **b** The function $\lambda_\kappa(\mu_0, \tau)$ from Equation S11 with fixed $\tau =5.4321$ and varying $\mu_0$: 0.5, 1, 4, 40. The red line corresponds to $\mu_0$=10 for which $\tau = 5.4321$ is critical, as described by Equation S13. **c** The function $\lambda_\kappa(\mu_0, \tau)$ from Equation S11 at the critical value $\tau_c$ defined in Equation S13 for different $\mu_0$: 0.5(green), 2(red), 5(yellow), 10(blue). If $\mu_0 < 1$ the maximum of $\lambda_\kappa$ is in $\kappa_m = 0$. As $\mu_0$ grows, the maximum approaches the limiting value of $\kappa_m = 1$. **d** The function $\lambda_{\tilde{\kappa}(\mu_0,\tau)}$ from Equation S26 with fixed $\mu_0$ =0.65 and varying $\tau$: 1,5,20,100. The red line corresponds to $\tau_c = 46.03$, as described by Equation S28. **e** The function $\lambda_{\tilde{\kappa}(\mu_0,\tau)}$ from Equation S26 with fixed $\tau =46.03$ and varying $\mu_0$: 1/4, 1/3, 1/2, 1. The red line corresponds to $\mu_0 = 0.65$, for which $\tau = 46.03$ is critical (Equation S28). **f** The function $\lambda_{\tilde{\kappa}(\mu_0,\tau)}$ from Equation S26 at the critical value $\tau_c$ defined in Equation S28 for different $\mu_0$: 0.25(green), 0.5(red), 1(yellow), 1.5(blue). If $\mu_0 < 1/3$ the maximum of $\lambda_\kappa$ is in $\kappa_m = 0$. As $\mu_0$ grows, the maximum $\kappa_m$ increases indefinitely.

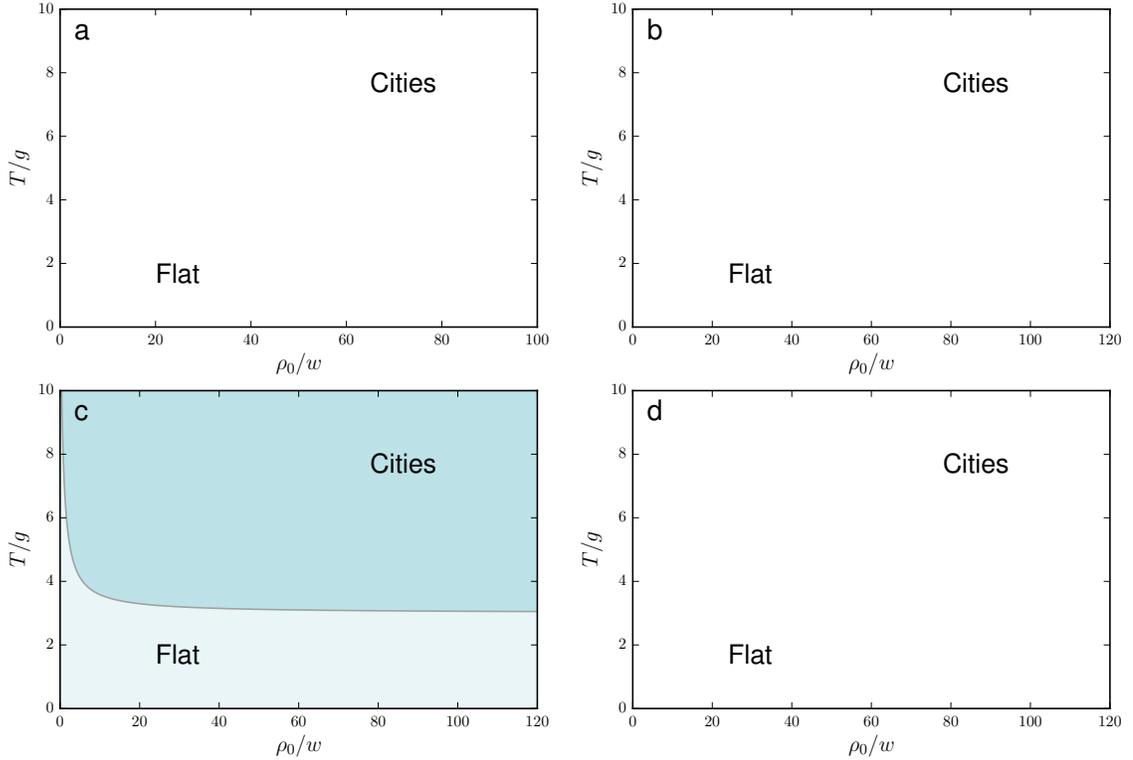

Figure 2: **a** Gravity Model - exponential f(r): Rescaled parameter space ($\mu_0$, $\tau$) described by Equation S13. Above the critical curve the formation of cities is possible whilst below it, any perturbations to the steady state distribution will decay back to $\rho_0$ and the landscape will be flat. **b** Gravity Model - power law f(r): Rescaled parameter space ($\mu_0$, $\tau$) described by Equation S13. By comparison with **a** we demonstrate that for the Gravity model the particular deterrence function has no effect on the condition for cities to form. **c** Intervening Opportunities Model: Rescaled parameter space ($\mu_0$, $\tau$) described by Equation S28. From comparison with **a** and **b** it can be seen that for an Intervening Opportunities model cities may form when the ratio between the migration rate and growth rate is lower than that required for Gravity models. **d** Radiation Model: Rescaled parameter space ($\mu_0$, $\tau$). By comparison with **c** it can be seen that if people relocate according to a Radiation model, the ratio between the migration rate and growth rate must be higher than that for an Intervening Opportunities model in order for cities to form.

20

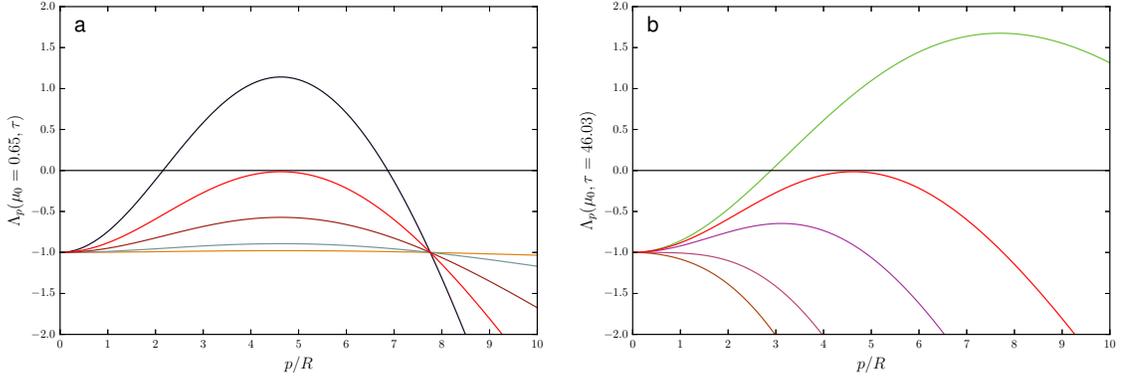

Figure 3: **a** The function $\lambda_{\tilde{p}}(\mu_0, \tau)$ for fixed $\mu_0 = 0.65$ and varied $\tau = 1,5,20,100$. The red line corresponds to $\tau_c = 46.03$, the same as the critical value for the 1D case. **b** The function $\lambda_{\tilde{p}}(\mu_0, \tau)$ for fixed $\tau = 46.03$ and varied $\mu_0 = 1/4,\ 1/3, 1/2, 1$. The red line corresponds $\mu_0 = 0.65$, for which $\tau = 46.03$ is critical.

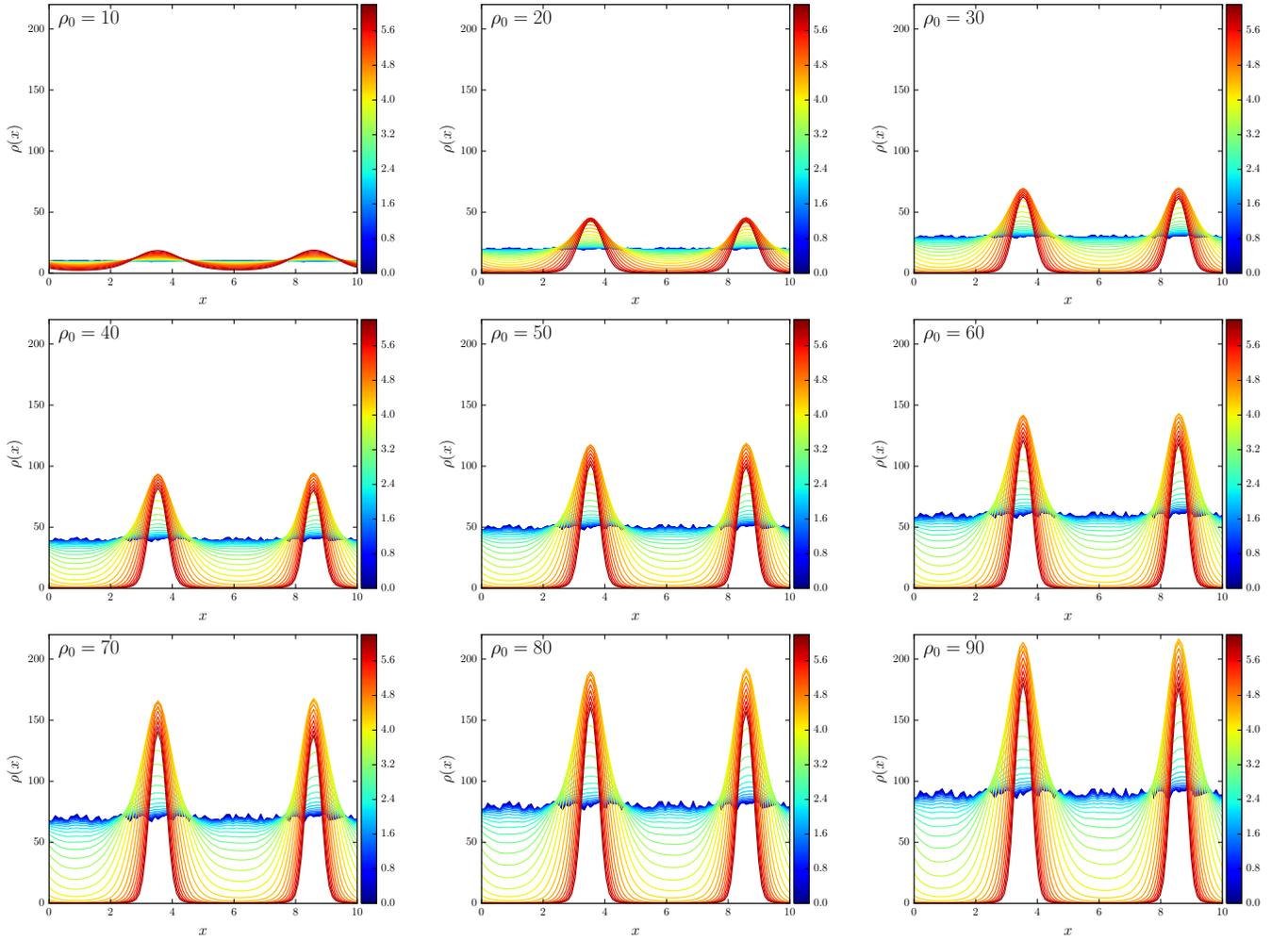

Figure 4: Gravity Model: Increasing $\rho_0$, fixed parameters: $T=10$, $R=1.5$.

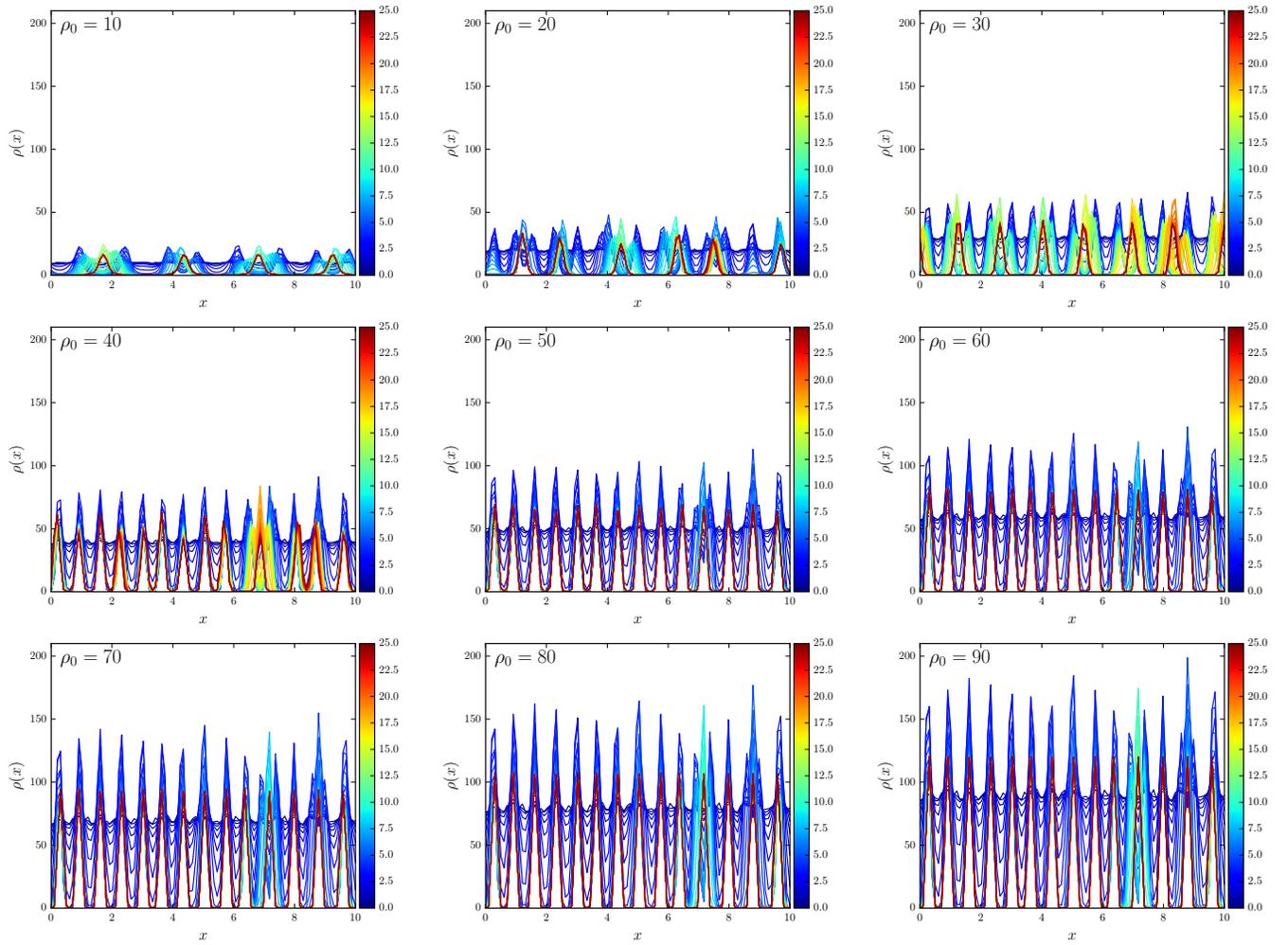

Figure 5: Intervening Opportunities Model: Increasing $\rho_0$, fixed parameters: $T$=10, $R$=0.5.

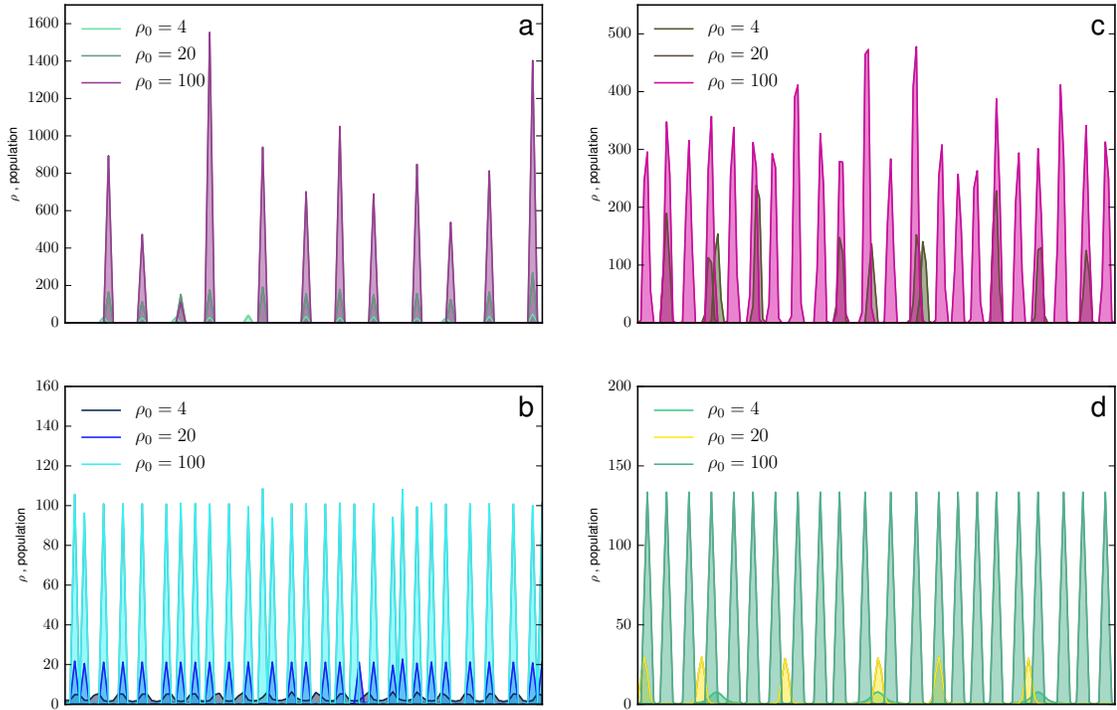

Figure 6: The final distribution obtained from **a** Gravity model, with parameters $T$=10, $R$=1.5, g=0 **b** Gravity model, with parameters $T$=10, $R$=1.5, g=1 **c** Intervening Opportunities model, with parameters $T$=10, $R$=0.1, g=0 **d** Intervening Opportunities model, with parameters $T$=10, $R$=0.1, g=1. $\rho_0$ is varied between 4 and 100 to demonstrate the relationship between the number of peaks and the steady state population density for each model.

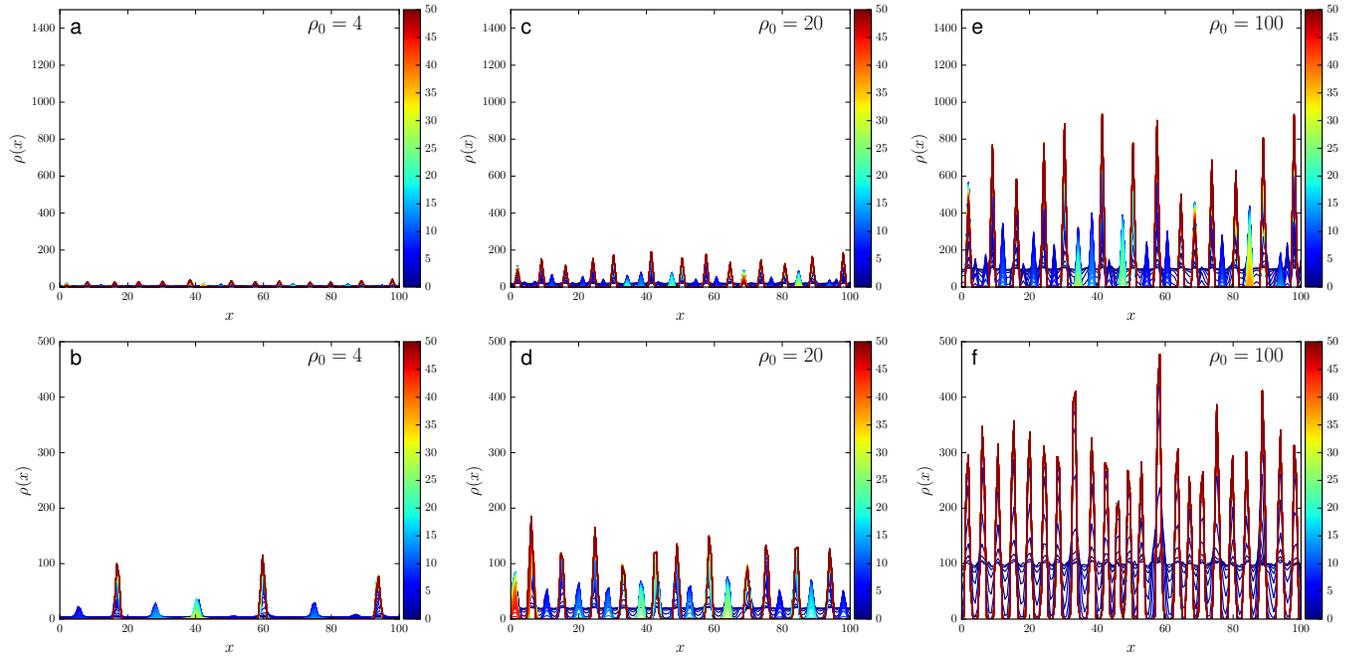

Figure 7: **a, c, e** Gravity Model: Increasing $\rho_0$ over the range 4, 20, 100 with fixed parameters $T$=10, $R$=1.5, g=0. It can be seen that the distributions are identical; the effect of increasing $\rho_0$ increases the height of the peaks only. The corresponding final states are displayed in Figure 6a. **b, d, f** Intervening Opportunities Model: Increasing $\rho_0$ over the range 4, 20, 100 with fixed parameters $T$=10, $R$=0.1, g=0. It can be seen that the effect of increasing $\rho_0$ is to increase the density of peaks along with the maximum height; the increase in height is not as great as that for the Gravity model as the total population is conserved. The corresponding final states are displayed in Figure 6c.

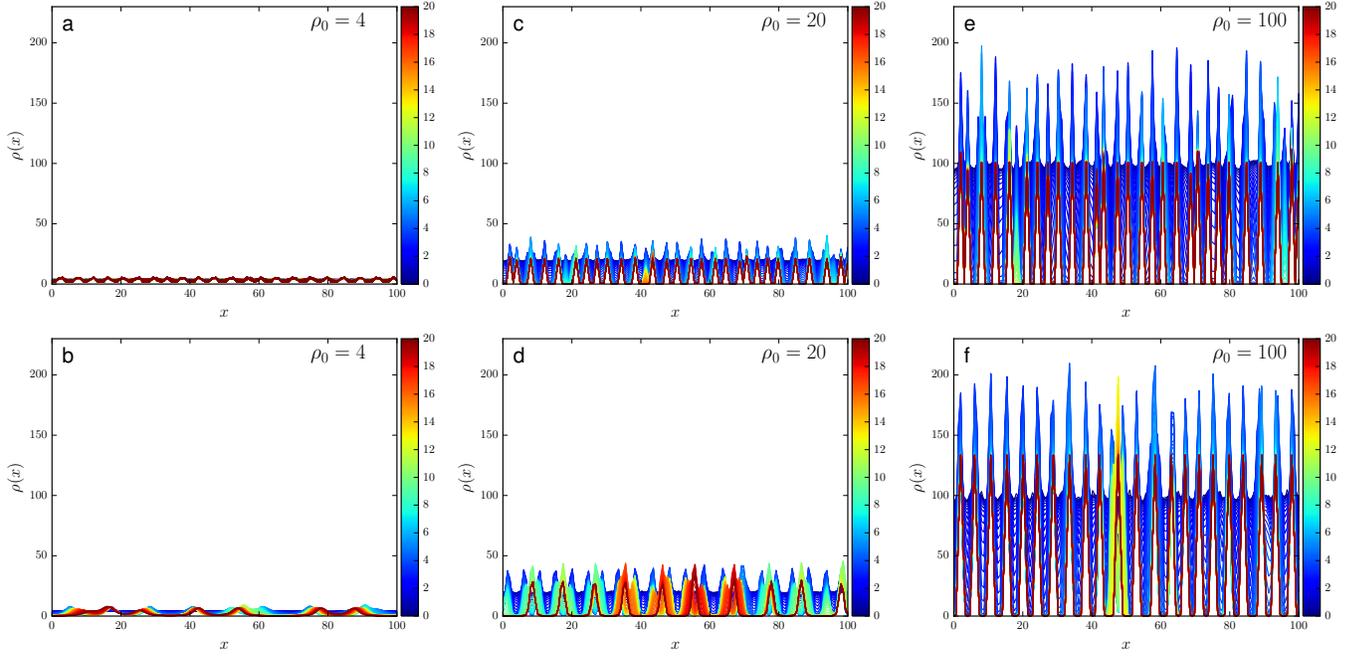

Figure 8: **a, c, e** Gravity Model: Increasing $\rho_0$ over the range 4, 20, 100 with fixed parameters $T$=10, $R$=1.5, g=1. It can be seen that the distributions are identical; the effect of increasing $\rho_0$ increases the height of the peaks only. The corresponding final states are displayed in Figure 6b. **b, d, f** Intervening Opportunities Model: Increasing $\rho_0$ over the range 4, 20, 100 with fixed parameters $T$=10, $R$=0.1, g=1. It can be seen that the effect of increasing $\rho_0$ is to increase the density of peaks along with the maximum height. The corresponding final states are displayed in Figure 6d.

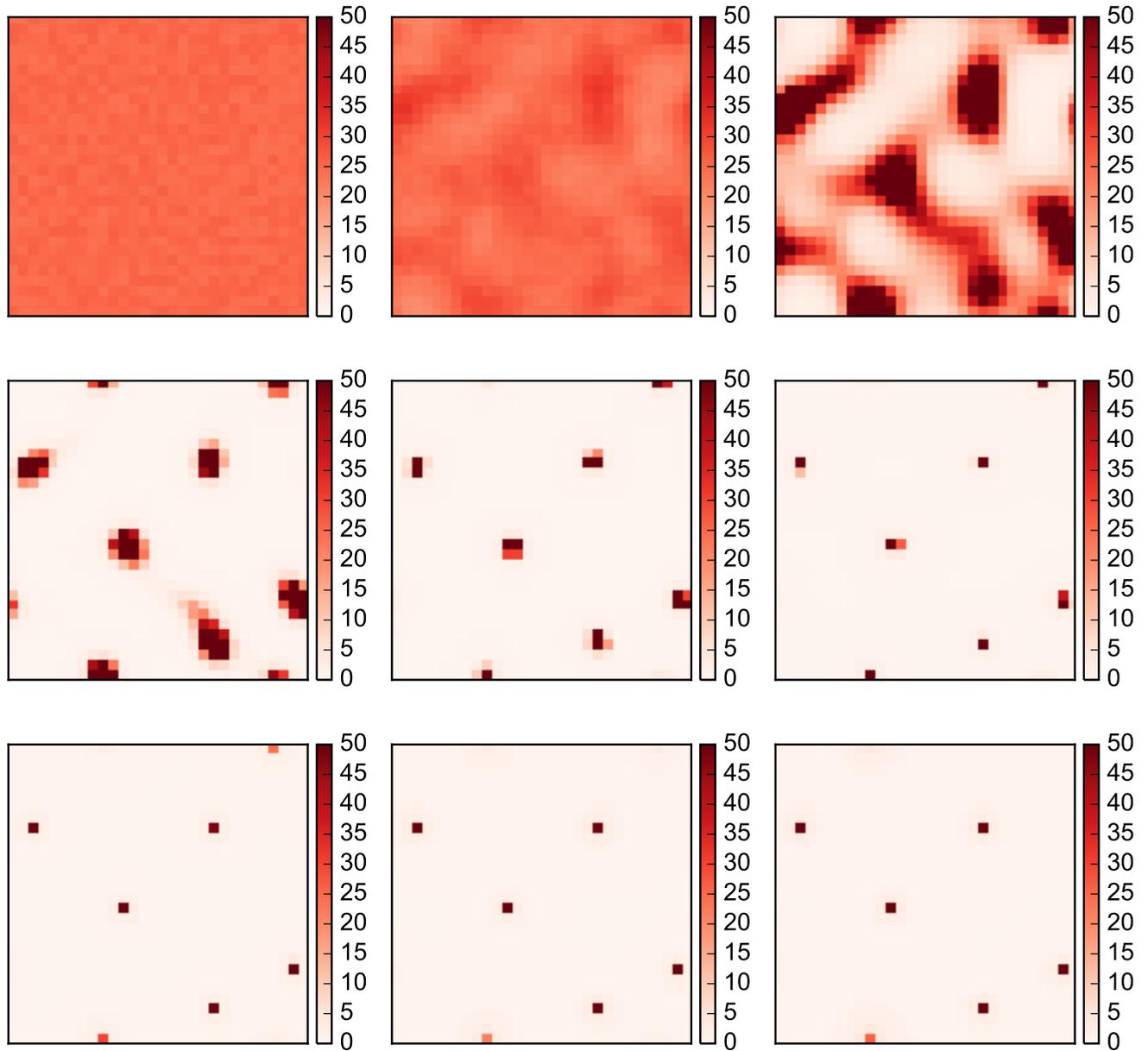

Figure 9: Gravity model: time evolution of an initial perturbation about $\rho_0 = 25$ with fixed parameters $L = 15$, $T = 30$ and $R = 0.5$. The spatial distribution of population converges to a stationary state of 6 cities.

26

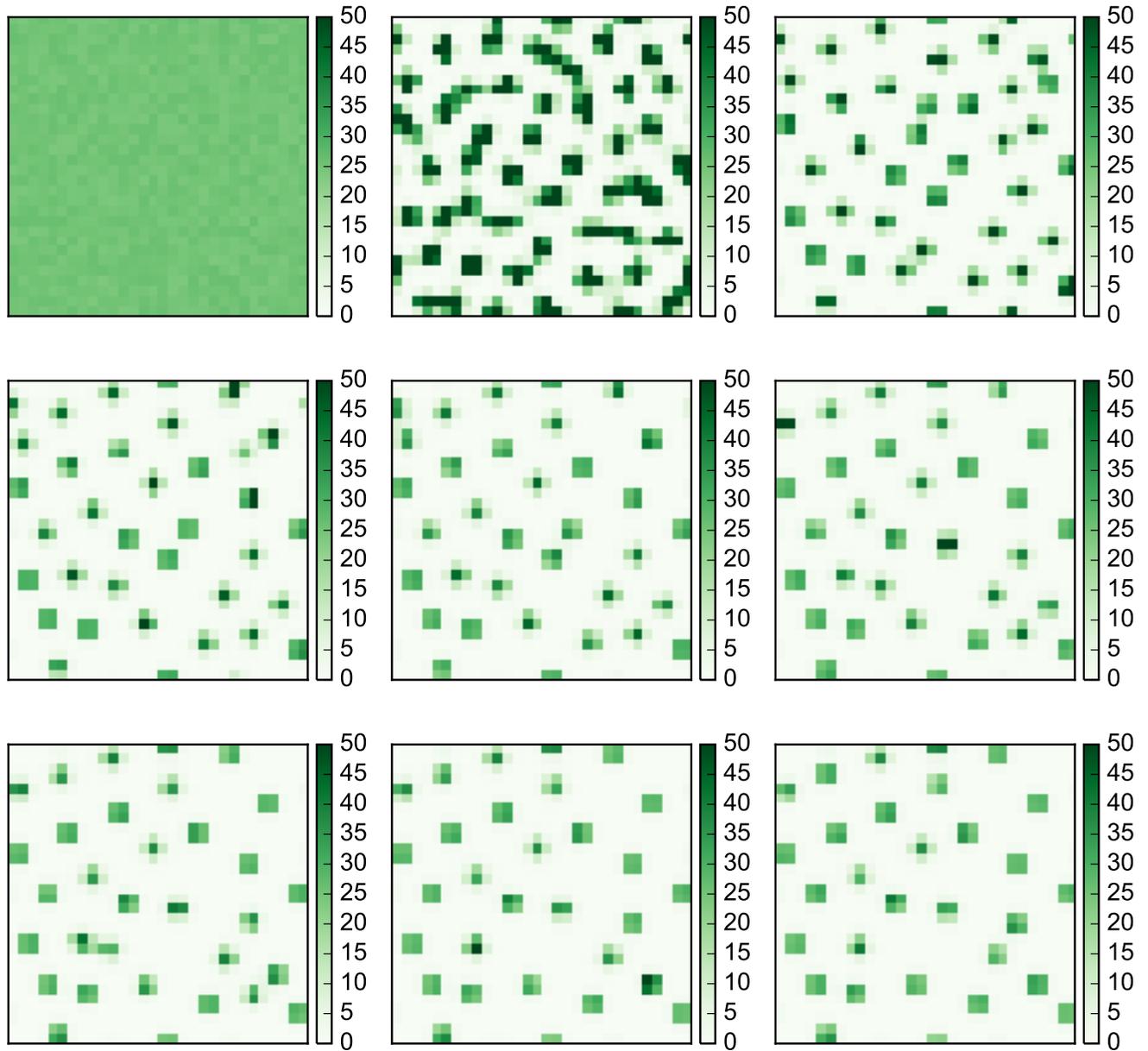

Figure 10: Intervening Opportunities model: time evolution of an initial perturbation about $\rho_0 = 25$ with fixed parameters $L = 15$, $T = 30$ and $R = 0.2$. The spatial distribution of population converges to a stationary state of 30 cities.

27

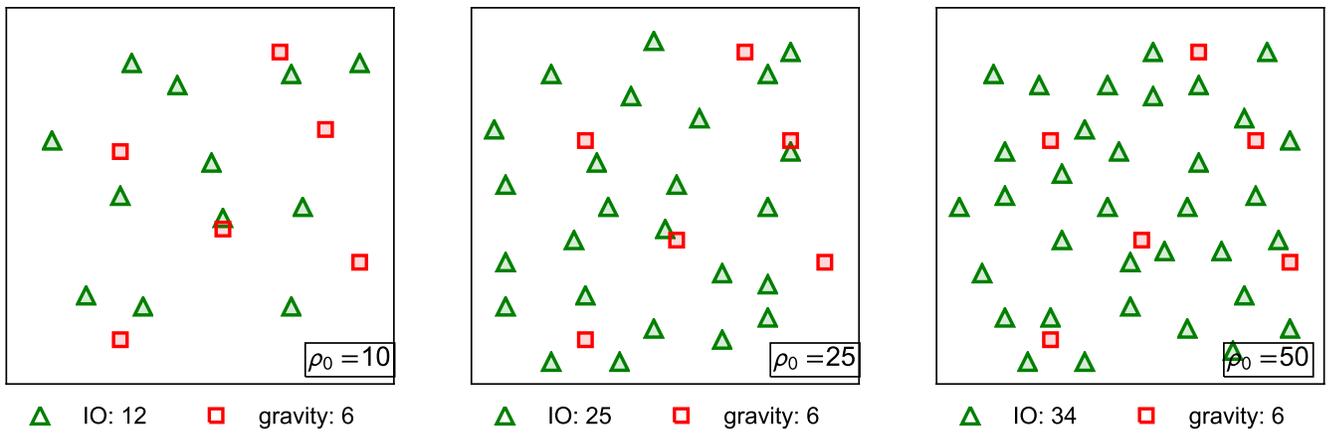

Figure 11: The final stationary distribution of cities when individuals relocate according to a Gravity model (red) and an Intervening Opportunities model (green) for increasing values of $\rho_0 = 10, 25, 50$ with fixed parameters $L = 15$, $T = 30$, $R = 0.5$ (Gravity model) and $R = 0.2$ (Intervening Opportunities model). It is observed that the final number of cities remains constant with increasing $\rho_0$ if individuals relocate according to a Gravity model whereas the number of cities increases with $\rho_0$ if individuals relocate according to an Intervening Opportunities model.

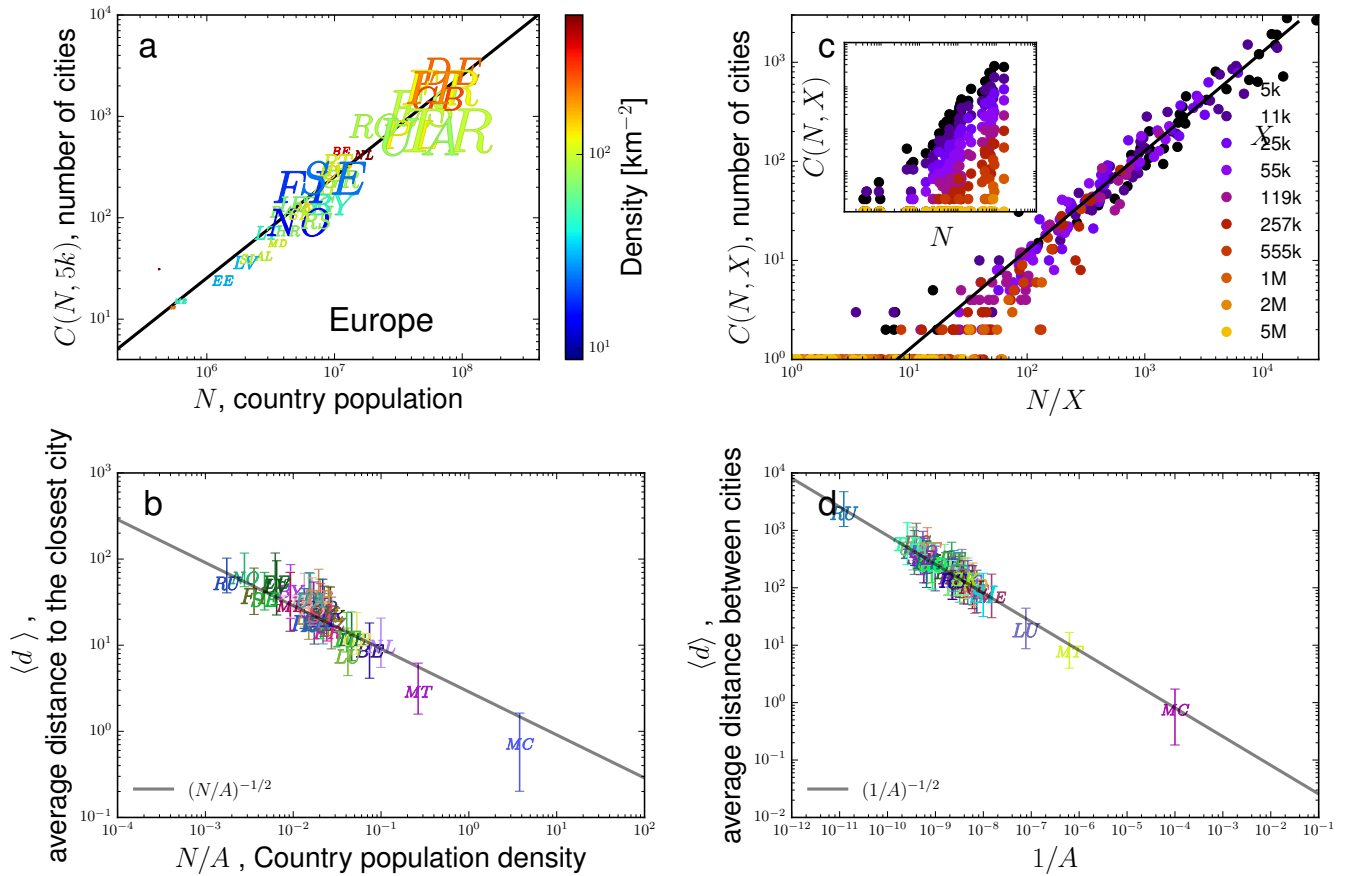

Figure 12: **a** Number of cities with population greater than 5,000 ($y$ axis) vs country population ($x$-axis), density (colour) and area (marker size) for countries in Europe (excluding Russia). The strong linear correlation between the number of cities and the country population supports the first law of urbanisation. **b** Number of cities with population larger than $X$ ($y$ axis) vs the ratio $N/X$ ($x$ axis); the ratio of the total population to minimum city population for countries in Europe. $X$ ranges from 5,000 to 5,000,000; as all points collapse on a single line we can conclude that the first law holds over several magnitudes of $X$. **c** The average distance to the closest city with population greater than 1,000 ($y$ axis) vs population density ($x$ axis) for countries in Europe. **d** The average distance between cities with population greater than 1,000 ($y$ axis) vs the inverse of the area ($x$ axis) for countries in Europe.

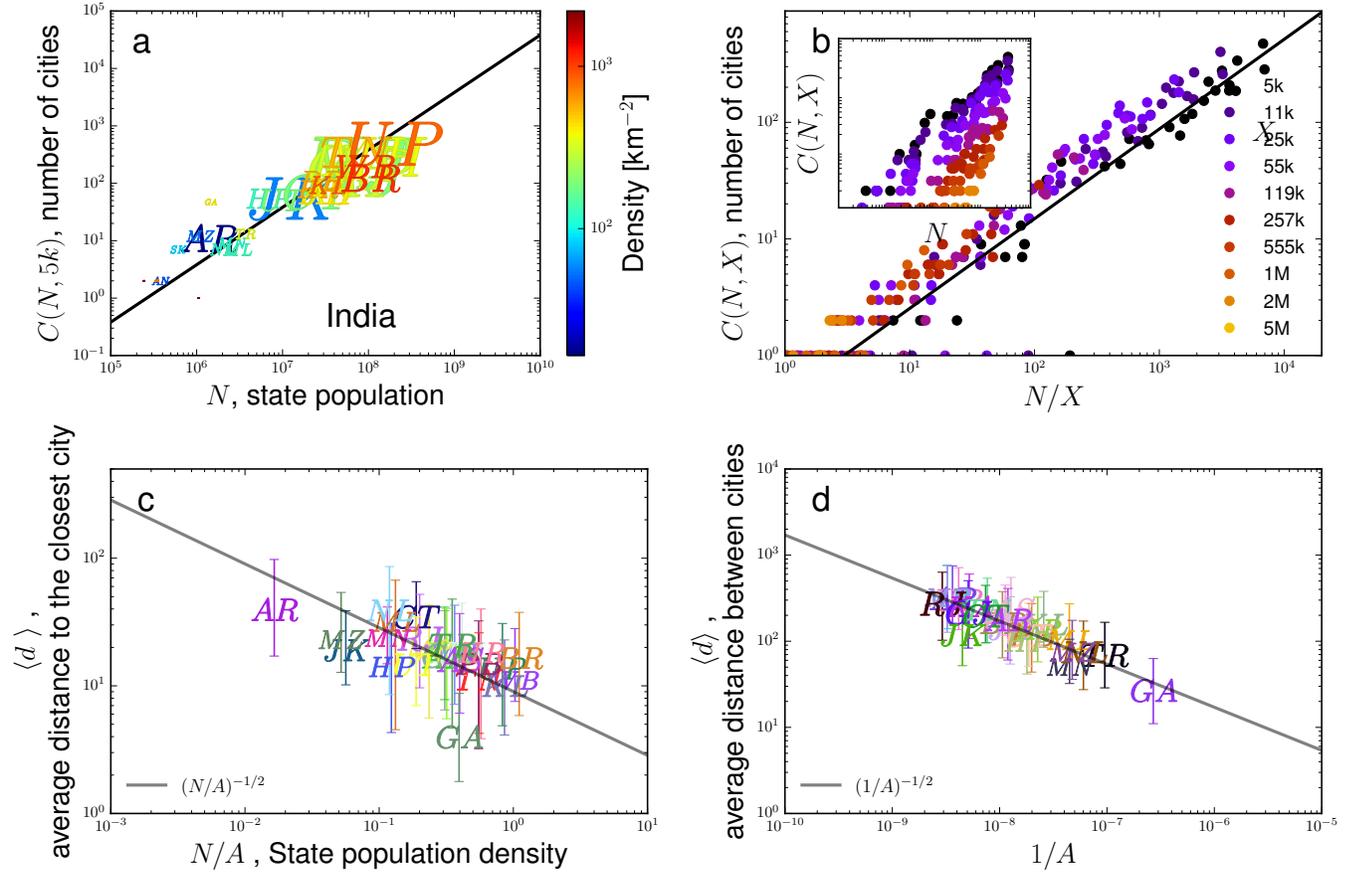

Figure 13: **a** Number of cities with population greater than 5,000 ($y$ axis) vs state population ($x$-axis), density (colour) and area (marker size) for states in India. The strong linear correlation between the number of cities and the state population supports the first law of urbanisation. **b** Number of cities with population larger than $X$ ($y$ axis) vs the ratio $N/X$ ($x$ axis); the ratio of the total population to minimum city population for states in India. $X$ ranges from 5,000 to 5,000,000; as all points collapse on a single line we can conclude that the first law holds over several magnitudes of $X$. **c** The average distance to the closest city with population greater than 1,000 ($y$ axis) vs population density ($x$ axis) for states in India. The state of Sikkim was removed from the data set as it only has two cities of population greater than 5,000. **d** The average distance between cities with population greater than 1,000 ($y$ axis) vs the inverse of the area ($x$ axis) for states in India.

30

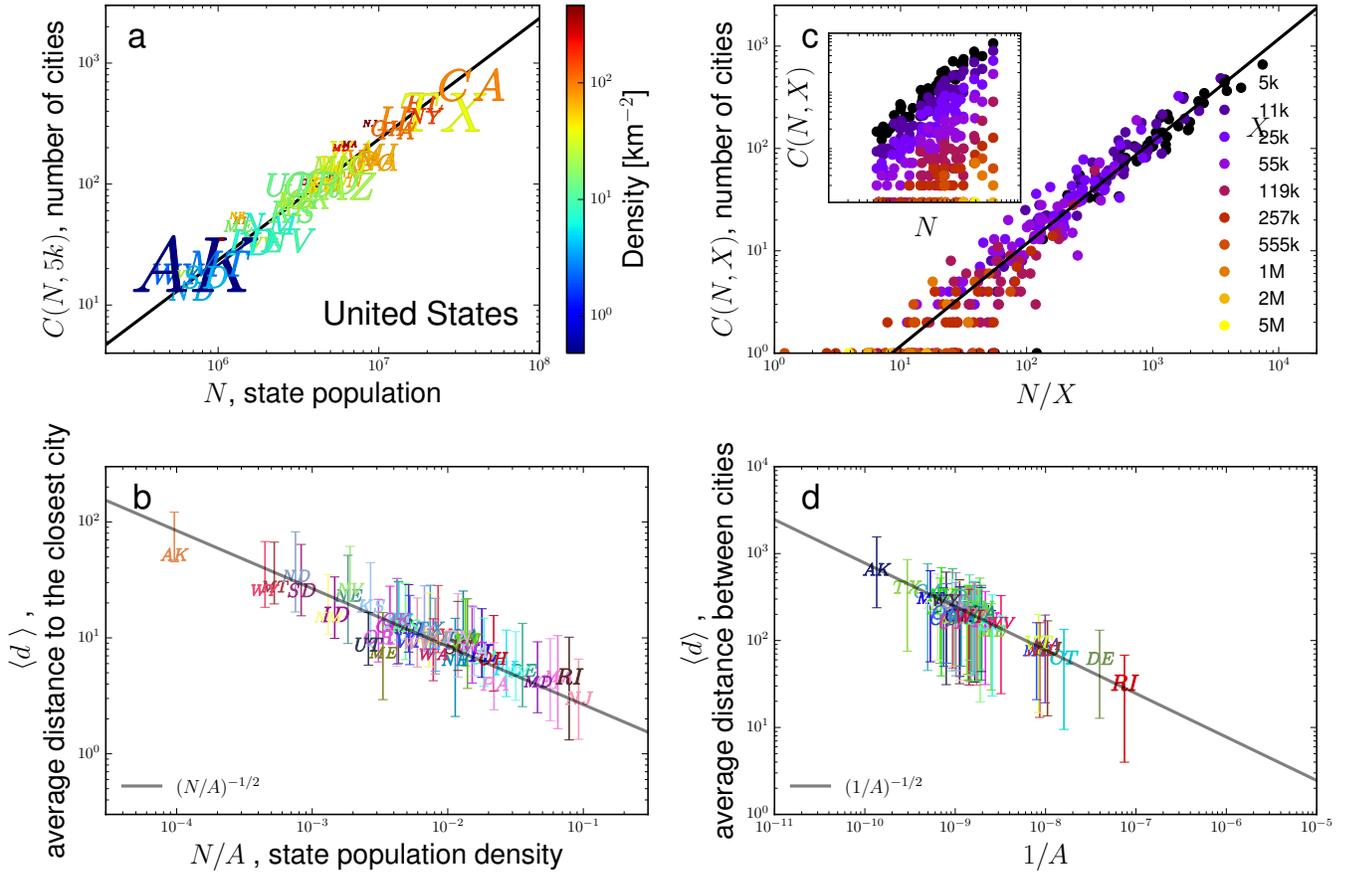

Figure 14: **a** Number of cities with population greater than 5,000 ($y$ axis) vs state population ($x$-axis), density (colour) and area (marker size) for states in the US. The strong linear correlation between the number of cities and the state population supports the first law of urbanisation. **b** Number of cities with population larger than $X$ ($y$ axis) vs the ratio $N/X$ ($x$ axis); the ratio of the total population to minimum city population for states in the US. $X$ ranges from 5,000 to 5,000,000; as all points collapse on a single line we can conclude that the first law holds over several magnitudes of $X$. **c** The average distance to the closest city with population greater than 1,000 ($y$ axis) vs population density ($x$ axis) for states in the US. **d** The average distance between cities with population greater than 1,000 ($y$ axis) vs the inverse of the area ($x$ axis) for states in the US.

31

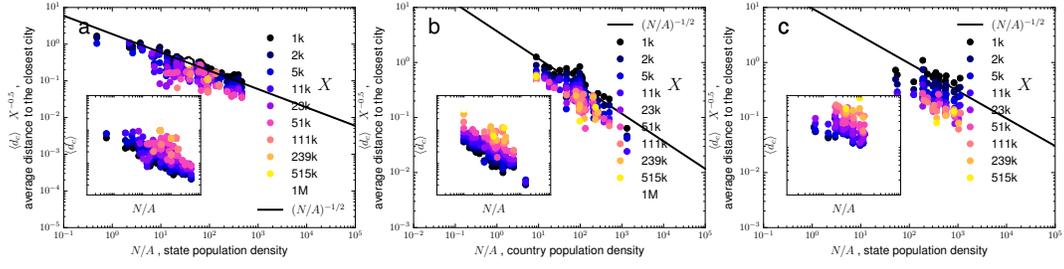

Figure 15: The average distance to the closest city, $\langle d_c \rangle X^{-0.5}$ ($y$ axis) vs population density $N/A$ ($x$ axis) for **a** states within the US, **b** countries within Europe and **c** states within India. $X$ ranges from 1,000 to 1,000,000; as all points collapse on a single line we can conclude that the second law holds over several magnitudes of $X$.

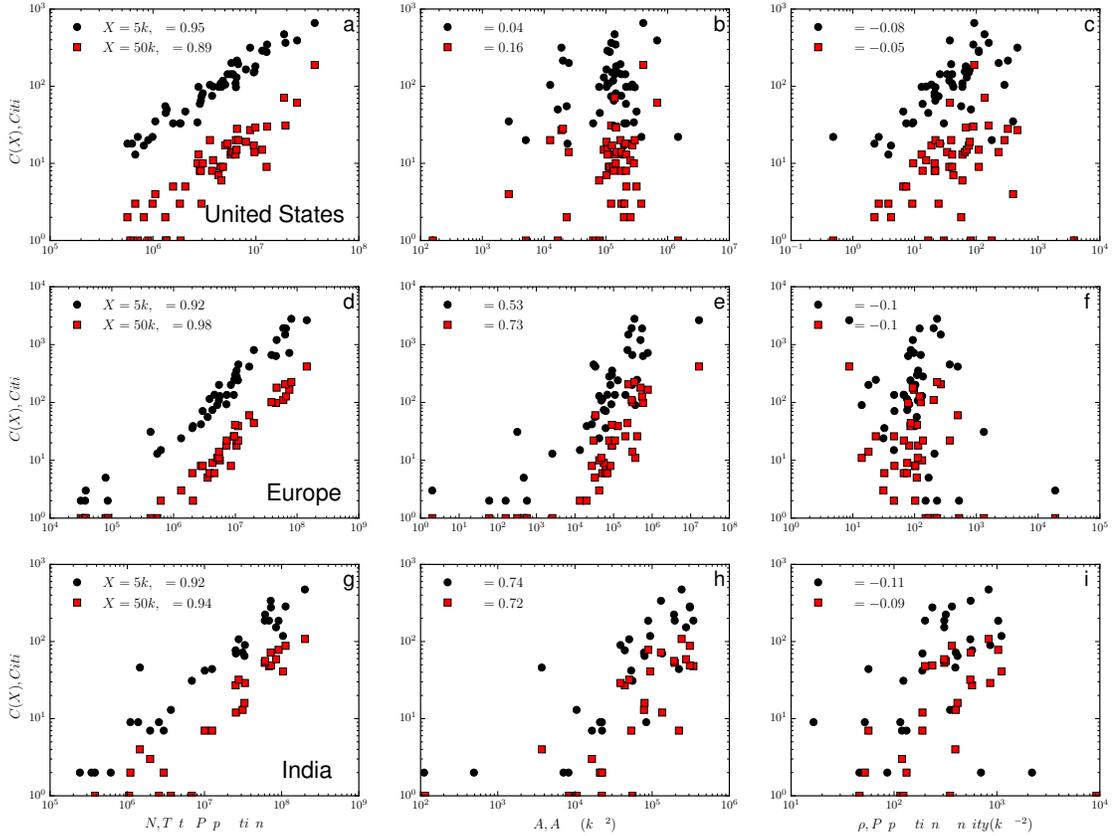

Figure 16: Correlation between total population $N$, area $A$, population density $\rho$ and the number of cities with more than $X = 5,000$ inhabitants (black) and $X = 50,000$ inhabitants (red) for **a-c** states within the US, **d-f** countries in Europe, **g-i** states within India.

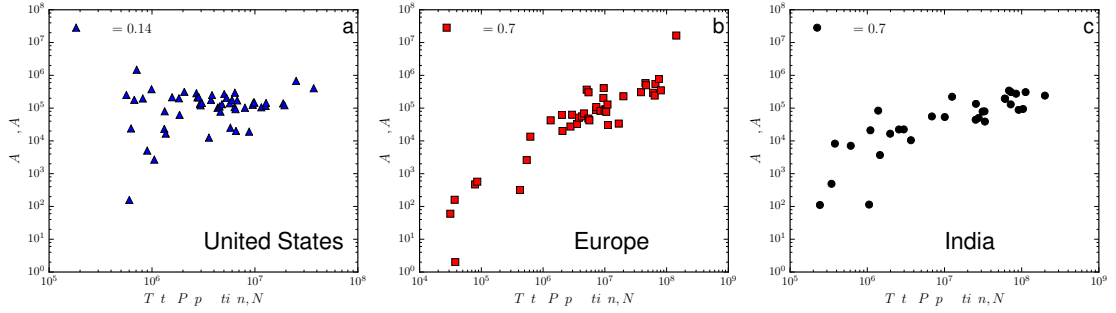

Figure 17: Correlation between total population, $N$, and area, $A$ for **a** states within the US, **b** countries within Europe, **c** states within India

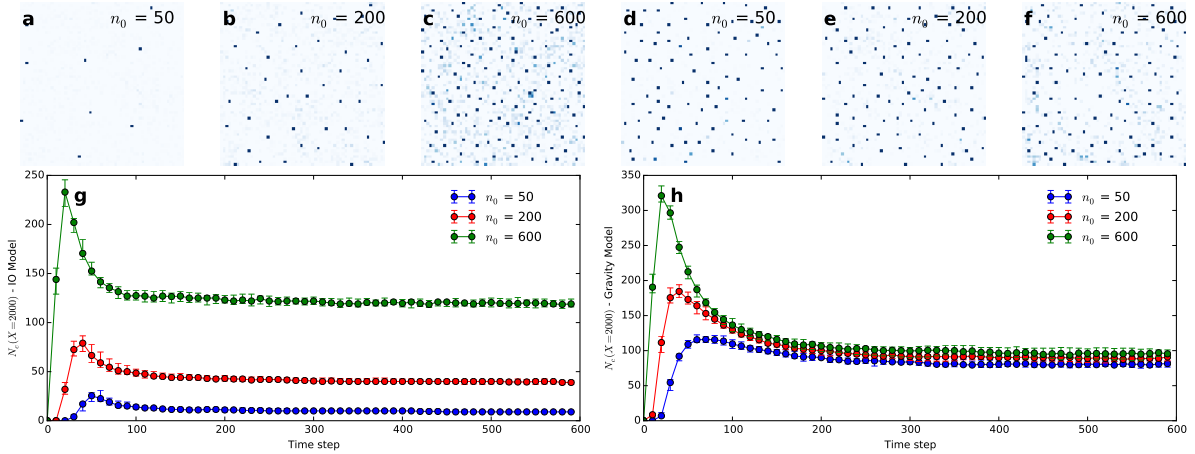

Figure 18: The final distribution of cities obtained from a set of stochastic simulations with population relocating according to **a-c** an Intervening Opportunities model and **d-f** a Gravity model with exponential deterrence functions and fixed probability of migration $T = 0.4$ and increasing initial population $n_0$ and parameters $X = 2000$, $\mu = -0.01$, $\sigma = 0.25$, $L = 60$ and $N = 60$. The evolution of the average number of clusters in each system, computed from 16 realisations of each set of parameters, is displayed for **g** an Intervening Opportunities model and **h** a Gravity model. It is observed that the average (median) number of clusters increases with $n_0$ when individuals relocate according to an Intervening Opportunities model whilst it tends to the same value for all $n_0$ when individuals relocate according to a Gravity model. Error bars correspond to the 10th and 90th percentile.

33

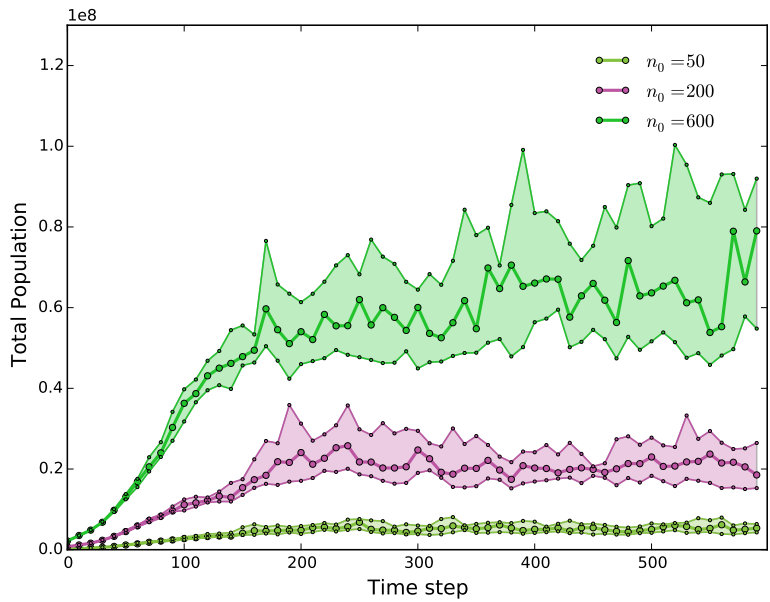

Figure 19: Total population vs time - Gravity model with and exponential deterrence function, increasing $n_0 = 50, 200, 600$ and fixed parameters $R = 1.5$, $T = 0.4$, $X = 2000$, $\mu = -0.01$, $\sigma = 0.25$, $L = 60$ and $N = 60$. The central lines within each region correspond to the mean total population at each time step where as the shaded area represents the region between the 25th and 75th percentiles.



\end{document}